\UseRawInputEncoding
\documentclass[english, spanish, catalan]{paper}

\usepackage{natbib}

\usepackage{graphicx}	
\usepackage{amsmath}	
\usepackage{amssymb}	
\usepackage{appendix}
\usepackage{float}
\usepackage{soul}

\usepackage{subfigure}
\usepackage{color}
\usepackage{enumerate}
\usepackage{wasysym}
\usepackage[section]{placeins}

\def\be{\begin{equation}}
\def\en{\end{equation}}

\title{Relativistic positioning: including the influence of the gravitational action of the Sun and the Moon and the Earth's oblateness on Galileo satellites}
\date{2020\\ November}

\author{\large Neus Puchades Colmenero  \\ \normalsize 
Institut Interuniversitari de Matem\`atica Multidisciplin\`aria. \\ \normalsize
Universitat Polit\`ecnica de Val\`encia. 
Cam\'i­ de Vera, S/N. Val\`encia 46022. And \\
\normalsize Florida Universit\`aria.  \`Area d'Enginyeria. \\ \normalsize 
Carrer del Rei En Jaume I, 2.
Catarroja, Val\`encia. 46470. Spain. 
\\ \large
Jos\'e Vicente Arnau C\'ordoba \\ \normalsize
Departament de Matem\`atica Aplicada. Universitat de Val\`encia. \\ \normalsize
Av. Vicent Andr\'es Estell\'es, S/N 
Burjassot 46100, Val\`encia, Spain. \\ \large
M\`arius Josep Fullana i Alfonso$^*$ \\ \normalsize
Institut Interuniversitari de Matem\`atica Multidisciplin\`aria. \\ \normalsize
Universitat Polit\`ecnica de Val\`encia.  
Cam\'i­ de Vera, S/N. Val\`encia 46022.}

\begin{document}


\maketitle






\begin{abstract}\label{abstract}
Uncertainties
in the satellite world lines lead to dominant positioning
errors. In the present work, using the approach presented in  \cite{neu14}, a new analysis of these
errors is developed inside a great region surrounding Earth.
This analysis is performed in the framework of the so-called
Relativistic Positioning Systems (RPS). Schwarzschild metric 
is used to describe the satellite orbits corresponding to the Galileo Satellites Constellation. 
Those orbits are circular with the Earth as their centre. They are defined as the nominal orbits.
The satellite orbits are not circular due to the
perturbations they have and to achieve a more realistic description such 
perturbations need to be taken into account. 
In \cite{neu14} perturbations of the nominal orbits were statistically
simulated. 
Using the formula from \cite{col10a} a user location is determined with the four satellites proper
times that the user receives and with the satellite world lines. This formula can
be used with any satellite description, although 
photons need to travel in a Minkowskian space-time. 
For our purposes, the computation of the photon geodesics in Minkowski space-time is 
sufficient as demonstrated in \cite{neu16}. 
The difference of the user position determined
with the nominal and the perturbed satellite orbits is computed. 
This difference is defined as the U-error. 
Now we compute the perturbed orbits of the satellites 
considering a metric that takes into account
the gravitational effects of the Earth, the Moon and the Sun and also the Earth oblateness.  
A study of the satellite orbits in this new metric is first introduced. Then
we compute the U-errors comparing the positions given
with the Schwarzschild metric and the metric introduced here.
A Runge-Kutta method is used to solve the satellite geodesic equations.
Some improvements in the computation of the U-errors using
both metrics are introduced with respect to our previous works. 
Conclusions and perspectives are also presented.
\end{abstract}

{\bf Keywords} Relativistic positioning systems.  Methods:
numerical.  Reference systems.

$^*$ {\it Corresponding author e-mail:} mfullana@mat.upv.es


\section{Introduction}\label{Intro}
%

In this work, the numerical codes developed in \cite{neu14} are used to compute
the positioning errors due to uncertainties in the satellite world lines, which are defined as the U-errors. In 
\cite{neu14} the U-errors were computed as the difference in a user positioning 
when describing the satellite world lines with Schwarzschild metric and a statistical perturbation
of those world lines. Now the description of the perturbed 
satellite world lines is made using a metric that takes into account
the gravitational effects of the Earth, the Moon, the Sun, as well as the Earth oblateness
(Earth quadrupole effect). 
This analysis is performed in the framework of the so-called
Relativistic Positioning Systems (RPS).
Both the satellite world lines and the user location are numerically computed  and then the corresponding U-errors are calculated.
Part of the work shown here was presented in \cite{Full19}.

Firstly, a short introduction on the difference between traditional positioning with Global Navigation Satellite Systems (GNSS) 
and positioning with RPS is presented.
Basically, for positioning, four satellites are needed to locate a user in space-time; i.e. a satellite for each unknown.
The system of algebraic equations that has to be solved to calculate the user location is given 
by Eqs.~(\ref{cap1-7}) (see section \ref{RPS} for more detail).
These equations are based on the null line element travelled by the photon from the emission to the reception event (see \cite{jua09}). 
The system given by Eqs.~(\ref{cap1-7}) has zero, one or two possible solutions.
A study of such solutions in RPS can be seen in \cite{neu12}. In the
present work we assume that only one solution is present.

The current GNSS are based on a Newtonian approach including relativistic corrections.
The GNSS would
work ideally if all the satellites and the receiver were at rest in an inertial reference frame.
Depending on the desired accuracy for the positioning, 
the necessary corrective terms have to be included
(see \cite{gom13}).
However, RPS are based on a
relativistic treatment from the beginning. For an account of this approach see \cite{col06a} and \cite{col06b}. 
Important work has been done using different relativistic models to describe the
motion of artificial satellites. \cite{Bru89} constructed the harmonic, 
dynamically non-rotating reference system for any body of the solar system 
and derived the equations of motion of test particles in the vicinity of the given
body using this reference system. This technique was applied to the Earth and its satellites. 
Meanwhile \cite{Dam94} introduced a particular General Relativity (GR) celestial mechanics framework. They computed
the equations needed for developing a complete relativistic theory of artificial Earth satellites. 
They claimed that their approach was more satisfactory than the previous ones especially 
with regard to its consistency, completeness and flexibility. 
In \cite{kos15} a model of RPS is presented in a
more realistic space-time near the
Earth with all important gravitational effects: Earth multipoles up to 
6th order, the Moon, the Sun, Jupiter, Venus, solid and ocean tides, and Kerr effect.
A recent approach was proposed by \cite{Roh16, Kyo18}. They implemented a full set of first-order Post-Newtonian
(PN) corrections in the high-fidelity orbit propagation KASIOP (Korea Astronomy and Space Science Institute
Orbit Propagator). And they numerically evaluated their effects on orbital elements for the
laser geodynamics satellite (LAGEOS) and laser relativity satellite (LARES) orbits. 
 
In the paper \cite{phi18} the relativistic orbital effects were estimated, which need to be considered for GRACE satellite orbits. 
They compared the magnitude of relativistic corrections in a GR space-time with PN corrections with various non-gravitational perturbations of satellite orbits 
and their results proved that for a GRACE satellite orbit (low Earth orbit), the relativistic acceleration is of the same order of several  
environmental effects. Hence, relativistic effects need to be considered for high precision space missions. The altitude for GRACE 
is approx. $456 \ Km$ and for Galileo approx. $23222 \ Km$, 
then the order of magnitude of all orbit effects for Galileo is slightly different from that of GRACE.

{\it a) Connection with previous works.}

In this research, some results about RPS, by \cite{col10a}, are applied in order to locate a user in the Earth vicinity up to $10^{5}$ Km, 
using the numerical codes developed in \cite{neu14}. 

In the Schwarzschild case, the time-like geodesic equations for a satellite were numerically solved in \cite{del09}.
The authors
presented a preliminary study for a satellite that moves in the equatorial plane. 
In this case, they calculated the emission coordinates, from the inertial ones,
for a few users with the same 
spatial coordinates who receive the satellite signal at different times.
The same problem was studied
using two other methods to compare the accuracies and the computational time.
Moreover, in \cite{cad10},
methods were described, in the Schwarzschild space-time,
to find the emission coordinates from the inertial ones and vice-versa. In \cite{neu16}, 
some aspects of the method described in \cite{cad10}, to obtain
the inertial coordinates (positioning) from
the emission coordinates, were used; for
example, a first-order approximation was used, in the parameter $G M_{\oplus}/r$, of the
time transfer function (see \cite{tey08}). Moreover,
in \cite{neu16}, other aspects such as the use of the analytical solution of \cite{col10a}
at zero-order calculations (which are very efficient) were considered. Also in \cite{neu16}, the positioning
errors associated to the simplifying assumption that photons move
in Minkowski space-time (S-errors) were estimated and it was shown that
the approach based on the assumption that
the Earth's gravitational field produces negligible effects on photons may be
used in a large region surrounding Earth. 

{\it b) The present work.}

In this paper, the computation of a satellite world line is performed. This calculation is done
in a GR space-time metric (see \cite{IAU2000} for more detail). 
The metric taken into account considers the effect of
the Earth, the Moon, the Sun and the Earth oblateness. Solar radiation pressure and other non-gravitational
effects are not considered although some of them have 
important contributions (see for instance \cite{Roh16, Kyo18}). However, 
the order of magnitude of their contributions depends on the distance to the Earth. 
Notice that here linear perturbation theory is not considered, as \cite{kos15} do. 
A metric is used from the first step to describe the space-time. Then 
satellite geodesic equations are computed. Afterwards a Runge-Kutta algorithm is
implemented to solve the geodesics ordinary differential equations. Although we are aware that
other effects should be considered we have decided to start with the three effects
commented in this paper to implement our algorithm. Later on we 
will continue incorporating more PN contributions and other terms.
Great accuracy in the digits is needed when considering all PN contributions.
The terms considered in the present work are sufficient to test the algorithm and 
prepare it for our 
future purposes in RPS.
Our results show that the greatest effect comes
from the Earth quadrupole, the second one from Moon and the third one from Sun (at the height of Galileo and GPS satellites). 
Those satellite world lines are needed to position a user in RPS. 

The perturbations computed here using metrics 
improve our previous works based on statistical methods 
as: 1) A better description of the real satellite world lines is achieved. 2)The effect
of each perturbing contribution in the satellite world lines is studied. 3) Also the combination of 
two of the three terms in the metric is studied and the three of them together. So, 
the orbits of the satellites are described depending on the terms considered. 4) Therefore, the contribution
of each effect on the user's positioning can also be studied. 
5) The value of the U-errors is now smaller. 6) That means a more precise calculation of
the user's positioning. 

{\it c) Schedule of this paper.}

Firstly, in section \ref{pert_ef}, the satellite world lines 
for a Galileo satellite (fixed orbital radius) are computed using the metrics of this paper. 
A study of the behaviour of such orbits is made. 
Afterwards, in section \ref{Dif_hs}, the satellite world line (in a space-time given by a metric 
with the gravitational potential of the Sun or the Moon or with the Earth's oblateness
or all effects together) is computed for a satellite at different orbital radius from the geocenter. In section \ref{RPS}, the satellite geodesics recently 
calculated are used to obtain the user location in space-time with RPS. The positioning errors are also calculated. Different users located on different spherical surfaces are considered. 
These spheres are centred in the geocenter and take different radii (from Earth radius to $5 \times 10^{4} \ Km$).
In Appendix \ref{justification}, a study of the order of magnitude of the orbit effects taken
here is computed, that is to say, Moon and Sun gravitational potential and Earth oblateness. 
These are some of the greatest effects on the orbits. In Appendix B, a short description of
circular orbits of Galileo satellites is written. 
In Appendix \ref{EDOs}, the time-like geodesic equations (calculating the Christoffel symbols) 
are written from the metric that describes a space-time, with Moon gravitational potential or Sun gravitational potential or Earth oblateness or 
taking into account all effects together. Finally, in section \ref{conclusions}, the main conclusions and perspectives are commented. 

{\it d) Notation and general considerations.}

Some aspects about the notation of this article are now commented. The Greek indices take values from $0$ to $3$ and the Latin indices go from 
$1$ to $3$ and $x^{0} \equiv ct$, where $t$ is the time coordinate and $c$ the light velocity. Also, we use $G$ for the Gravitational constant and $M_{A}$ 
for the mass of the body $A$. The symbols (subscripts) $\oplus$, $\odot$ and $\leftmoon$ are assigned to Earth, Sun and Moon, respectively. 
We use this notation throughout the article. The symbol $AU$ is for one Astronomical Unit. The Minkowski tensor $\eta_{\mu \nu}$ is defined as: $\eta_{\mu \mu} \equiv (-1,1,1,1)$ 
for the diagonal elements and null for the rest of components. In our codes and in the formulae, we choose the units in such a way than the light velocity is $c=1$.

Moreover, throughout this paper, the reference system considered is the Geocenter Reference System ($GRS$).
An appropriate method is used to represent some quantities in 3D, $t=constant$ (user time), space-time sections.\\Colour bars and an appropriate pixelization are necessary. 
In this paper, as in \cite{neu12}, \cite{neu14} and \cite{neu16}, the HEALPIx ({\it hierarchical equal area isolatitude pixelisation of the sphere}) package (\cite{gor99}) is used, 
with the same parameters and projections than in the paper \cite{neu14}.
 
We simulate the world lines of the Galileo background configurations and we use these world lines as initial conditions for the ordinary differential equations (ODE). 
The Galileo constellation is composed by 27 satellites ($n_{s} = 27$), located in three equally spaced orbital
planes (9 uniformly distributed satellites in each plane).
The inclination of these planes is $\alpha_{in} = 56 \ deg$ and the
altitude of the circular orbits is $h = 23222 \ Km$; thus,
the orbital period is close to $14{.}2 \ h$. The satellites are
numerated in such a way that the satellites 1 to 9,
10 to 18, and 19 to 27 are placed in distinct consecutive
orbital planes. Initially, the trajectories are assumed to be
circumferences whose centres are located in the Earth
centre, which is also the origin of the almost inertial
reference system used for positioning. In sections \ref{pert_ef} and \ref{Dif_hs}, the Galileo satellite used is number $1$.

\FloatBarrier
\section{\bf Satellite world lines (Galileo Constellation)} \label{pert_ef}

This section shows the Galileo satellite world lines
computation with our algorithm. 
The description of the satellite world lines
with Schwarzschild metric gives
circumferences centred in the centre 
of the Earth. The Earth is approximated 
as spherically symmetric and with no rotation.
These trajectories are defined as nominal trajectories.
A short description of the 
circular satellite world lines can be 
seen in Appendix B. 
In the present work, the Schwarzschild satellite world lines are
perturbed to compute the effects of the Sun, the Moon and the Earth quadrupole. 
A detailed description of the
solution of the Ordinary Differential Equations (ODE)
to compute the satellite world lines in this metric can be seen in
Appendix \ref{EDOs}. 
The numerical solution of such ODE is also explained.
This section is divided in two subsections. 
The first one \ref{Tidal}, 
presents computations taking into account only
the Moon and Sun gravitational fields and their sum. The second one
\ref{J2} presents the effects of the Earth quadrupole on the orbits too. 
This second subsection includes the computations of
the effect on the satellite of the Earth oblateness alone  
and of this oblateness plus the sum of the effects of Sun and Earth.
In all computations, the alignment of Moon and Sun respect 
to the satellite has been taken into account when computing the orbit of the satellite.
Notice that the figures represent the deviation from the 
corresponding circular orbits, nominal orbits, that would follow the satellites if 
the Earth's field would describe 
them, Schwarzschild metric. In such figures, the 
circular orbits would be straight horizontal lines. 

Let us remark that this section and the next one are introduced so as to
clearly describe the perturbing effects that are considered
in this research. Once those effects have been presented, 
we use them to define the
U-errors in a similar way as it was performed
in \cite{neu14} but now with the new metrics considerations. 
That is to say, the perturbations are not 
computed with a statistical algorithm but
with some physical effects that affect the 
satellites. 
This represents an advance from the previous
work as it can describe more realistic cases.

\subsection{\bf Sun and Moon effects} \label{Tidal}
A gravitational term (Sun or Moon) can be considered as
a satellite orbit perturbation. 
The $B$ gravitational term reads:
\begin{equation}
 \phi^{B}=2G\left ( \frac{M_{B}}{r_{B}} \right )
\label{pertur_Moon_or_Sun}
\end{equation}
where $B$ is the celestial body ($B=\leftmoon , \ \odot$), either Sun or Moon (in our case).\\
The first spatial derivatives with respect to the additional $B$ potential stand:
\begin{equation}
 \phi_{,i}^{B}=-2G\left ( \frac{M_{B} \left (x^{i}-x_{B}^{i} \right )}{r_{B}^{3}} \right )
\label{dphi_Moon_or_Sun}
\end{equation}

The initial conditions for the Moon and Sun are given from the Ephemeris in the GRS 
(reference plane: equatorial and rectangular coordinates). 
The Miriade Ephemeris Generator is used. 
Therefore, the initial conditions (velocity and position vectors) for the Moon and the Sun are the corresponding Ephemeris for the time $2018-12-13T17:00:00.00$. 
See \cite{IAU2000} for more details about initial conditions. 
Nevertheless, the initial conditions for the satellite (velocity and position vectors) 
considered here
are a Galileo satellite moving in a circular orbit with angular, $\Omega$, 
and linear, $v$, velocity from $\Omega=\sqrt{\frac{G M_{\oplus}}{R^{3}}}$, $v=\sqrt{\frac{G M_{\oplus}}{R}}$. 
The following factor is also considered:
 
\begin{equation}
\label{gamma1}
\gamma=\frac{dx^{0}}{d\tau}
\end{equation}

with:

\begin{equation}
\label{gamma2}
\gamma=\frac{1}{\sqrt{1-\frac{3 G M_{\oplus}}{R}}} 
\end{equation}

Notice that the relativistic factor Eq.(\ref{gamma2}), enhances the GR treatment with the PN approximation (see Eq. (A.36), p. 210 of  \cite{Gou13}):


Fig.~\ref{fig_ESM_ephemerides} plots 
the radial distance, $R$, of a Galileo satellite, for two orbital periods, from the geocenter.
There are three different computations: a) Considering only the Moon effect (blue). b) Considering
only the Sun effect (red). c) Considering the Moon plus the Sun effect together (magenta).  
Moon and Sun contributions are considered in the metric. 
Fig.~\ref{fig_ESM_ephemerides} allows us to compute the variation of the distance of the
satellite to the geocenter, this indicates the deviation from the circular orbit. 
The circular orbit is the one described with the metric considering only the Earth
contribution: Schwarzschild metric. In the figure it would correspond to
a straight horizontal line that would start from the point where the 
other lines start from left to right. 
Therefore
the effect of each perturbation alone or combined on the satellite orbit can be measured. 
One way to compute this variation is to calculate the difference between two different positions of the 
satellite in its orbit. The maximum variation of the distance to the geocenter 
will be the difference between the farthest position and the nearest one. 
Notice that absolute values are computed. 
$R_{max}$ is this maximum distance (maximum in 
the plot) and $R_{min}$ is the minimum one (minimum in the plot)  
when two orbital periods are considered. 
So, the maximum 
variation of the radial distance of the satellite to the geocenter, 
when two orbital periods are considered is $R_{max}-R_{min}$. 

Next the values of this variation, $R_{max}-R_{min}$, 
for the cases we have considered are here presented
as it can be seen in Fig.~\ref{fig_ESM_ephemerides}: 
\begin{itemize}
\item Around $600 \ m$ when only the gravitational contribution from Moon is considered.
\item Around $200 \ m$ when only the gravitational contribution from Sun is considered.
\item Around $700 \ m$ when the gravitational contributions from Sun plus Moon are considered together.
\end{itemize}
The order of magnitude obtained is the one expected 
for every quantity computed in this paper as 
(see the book \cite{teu15}, Chapter 3, and \cite{Kyo18}): 
\begin{itemize}
\item At several hundreds of meters the second-largest effect producing 
orbit changes is the acceleration caused by Sun plus Moon. 
\item In spite of the fact that Sun is much more massive, the tidal
acceleration in the geocentric frame is bigger for the Moon.
This fact is due to the much lower distance of Moon to the satellite 
(according to Table \ref{table_perturb}, column 4, Appendix \ref{justification}).
\end{itemize}

\FloatBarrier
\subsection{\bf Adding the Earth oblateness} \label{J2}

The gravitational potential for Earth oblateness is (see \cite{mon05} and \cite{teu15}:
\begin{equation}
 \phi^{J_{2}}=-2G \left ( \frac{J_{2} M_{\oplus} R_{\oplus}^{2} P_{2}\left ( \cos \theta \right )}{r_{\oplus}^{3}}\right )
\label{pertur_J2}
\end{equation}
where the Legendre polynomial of degree 2 is 
$P_{2}\left ( \cos \theta \right )= (3 \cos^{2} \theta-1 )/2 $. 
The value of the zonal gravitational coefficient of degree 2 of the Earth $J_{2}$ is
$J_{2}= 1{.}08263 \times 10^{-3}$ and $\theta$ and $R_{\oplus}$ are the co-latitude of the satellite and the equatorial radius of the Earth, respectively.

The first spatial derivatives due to Earth oblateness are:
\begin{eqnarray}
\phi_{,i}^{J_{2}}=-3 G \left ( \frac{J_{2} M_{\oplus} \left (x^{i}-x_{\oplus}^{i} \right ) R_{\oplus}^{2} }{r_{\oplus}^{7}}\right ) \times \nonumber \\ 
\left[ A(i) \sum_{j=1}^{3}  \left (x^{j}-x_{\oplus}^{j} \right )^{2}-5\left (x^{3}-x_{\oplus}^{3} \right )^{2} \right] 
\label{dphi_J2}
\end{eqnarray}
where $A(i)$ is 1 for $i=1,2$ and  is 3 for $i=3$.

Fig.~\ref{fig_ESMJ2_ephemerides} plots radial distance, as a function of proper time 
(for 2 orbital periods), of a Galileo satellite from the geocenter (see previous subsection
for the definitions of the quantities used here). 
In this subsection, we only present the
effects considered in the metric of i) the Earth oblateness and 
ii) Moon plus Sun plus Earth oblateness.
The study of Fig.~\ref{fig_ESMJ2_ephemerides}, gives 
the maximum variation of the radial distance of the satellite, 
$R_{max} - R_{min}$ (as defined in the previous subsection), during two orbital periods:
\begin{itemize}
\item About $2 \ Km$ for Earth oblateness.
\item About $3 \ Km$ for Earth oblateness plus direct tides from Sun and Moon.
\end{itemize}

Again we obtain the expected order
of magnitude for the effects considered in this paper: 
the greatest effect on GNSS satellites is mainly due to
Earth's oblateness (see the book \cite{teu15} and \cite{Kyo18}). That is, 
effects at kilometre level are obtained.  

\begin{figure}
\center
\includegraphics[width=12cm, height=8cm]{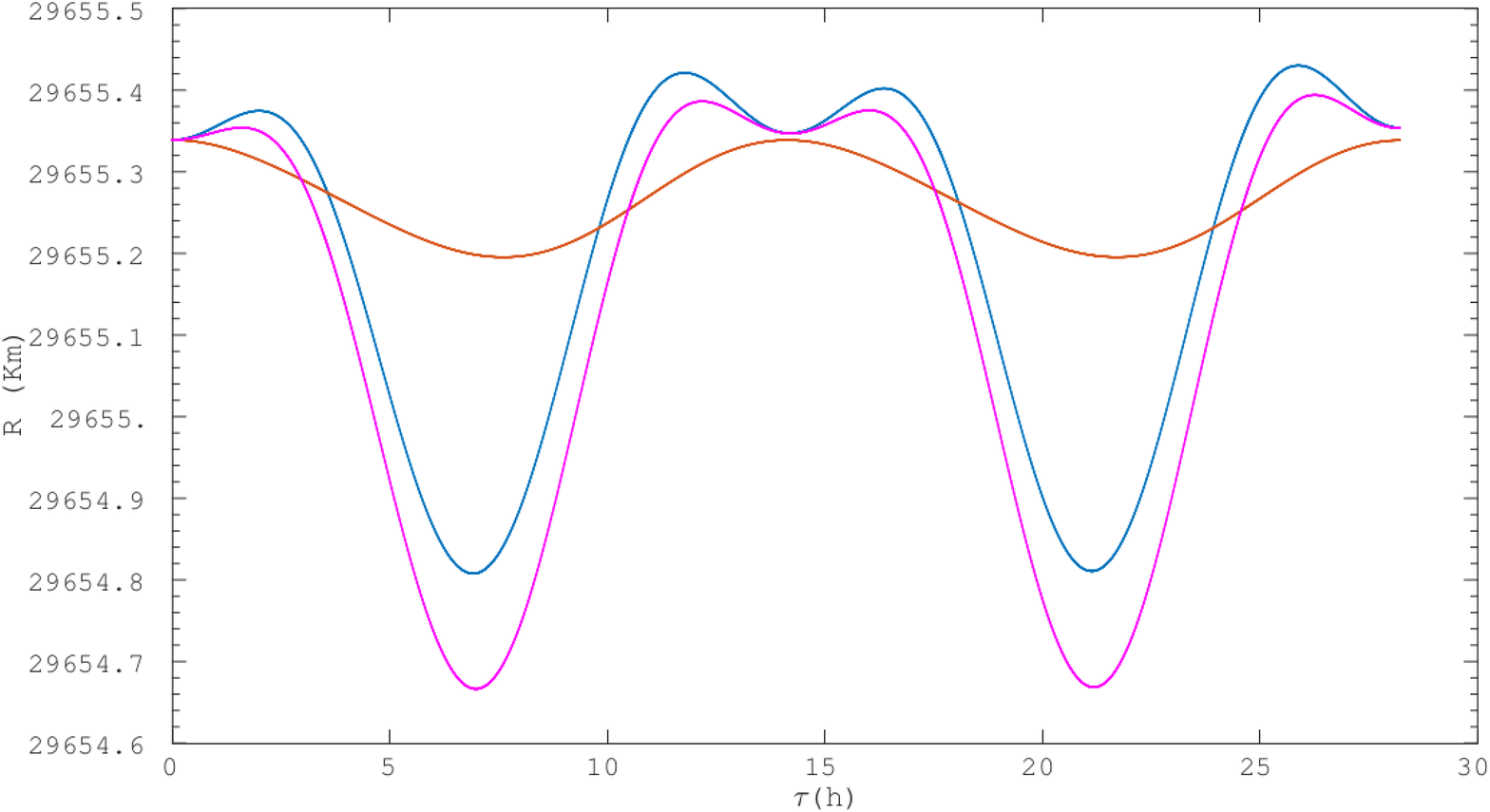}
\caption{Radial distance (in $Km$), versus the proper time (for two orbital periods), of a Galileo satellite from the geocenter. 
The gravitational contributions that are considered in the metric are Moon ($blue$), Sun ($red$) and Moon plus Sun ($magenta$).}
\label{fig_ESM_ephemerides}
\end{figure}

\begin{figure}
\center
\includegraphics[width=12cm, height=8cm]{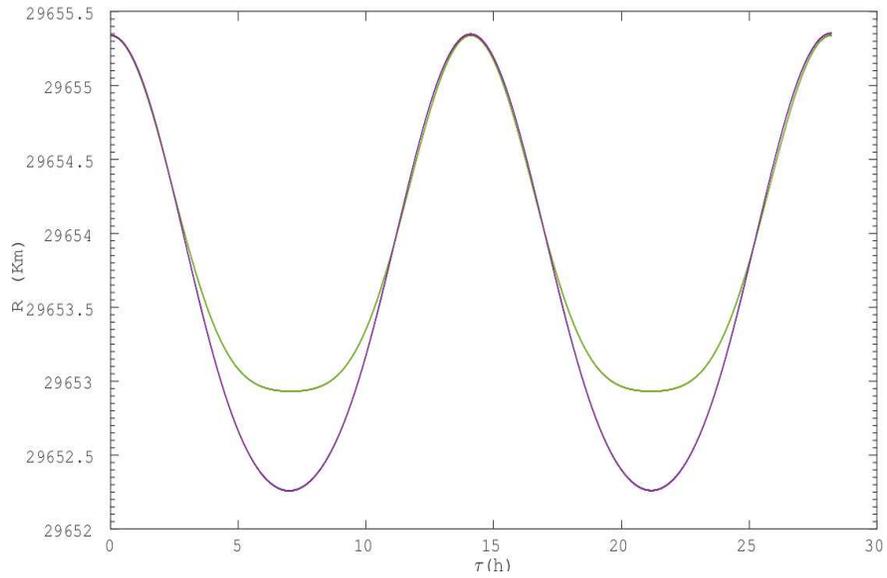}
\caption{Radial distance (in Km), as a function of the proper time 
(2 orbital periods), 
of a Galileo satellite from the geocenter. 
Earth oblateness ($green$) and Moon plus Sun effect plus Earth oblateness ($violet$) are here taken into account.}
\label{fig_ESMJ2_ephemerides}
\end{figure}

We now compare the results of Figs.~\ref{fig_ESM_ephemerides} and \ref{fig_ESMJ2_ephemerides}.
As it can be observed in
Figs.~\ref{fig_ESM_ephemerides} and \ref{fig_ESMJ2_ephemerides},
the recovering of the satellite radial distance in 1 or 2 orbital periods is achieved in the
following cases:
\begin{itemize}
\item Direct tides from Sun (see Fig.~\ref{fig_ESM_ephemerides}),
\item Earth oblateness (see Fig.~\ref{fig_ESMJ2_ephemerides}),
\end{itemize}
that is to say, when direct tides from Moon are not considered. Satellite position shifting is obtained when Moon is considered. 

Notice that the order of magnitude of the R-axes is different in both figures.
In Fig.~\ref{fig_ESM_ephemerides} the difference between each division is of 100 meters while in 
Fig~\ref{fig_ESMJ2_ephemerides} each difference is of 500 meters. This is done so because the
order of magnitudes of the effects are different, greater when the oblateness is introduced, as it can be seen in the plots.
Also notice that the presence of the Moon causes the bump in the effects from Moon + Sun 
(magenta in Fig.~\ref{fig_ESM_ephemerides}). The reason for not observing this bump in  
Fig.~\ref{fig_ESMJ2_ephemerides} in the
Moon + Sun + Earth oblateness (violet in Fig.~\ref{fig_ESMJ2_ephemerides}, where 
Moon effect is also considered) is this difference in the order of magnitude of
divisions of R-axes. We are more interested in emphasizing orders of magnitude and
that's why this bump is smoothed.


\FloatBarrier
\section{Orbital perturbation effects for different orbital radius from the geocenter}\label{Dif_hs}

In this section, the satellite world 
lines are calculated with our algorithm for different 
satellite orbital radii from the geocenter.
So we extend the computation to other possible satellite orbits far from Galileo ones. 
The orbital perturbation effects are estimated for these satellite orbital radii.
Again, the radial distance $R$ versus proper time is evaluated for different satellite orbital 
perturbating effects: presence of Moon and Sun gravitational field and Earth quadrupole.

\subsection{Considering different orbital perturbation effects for a given satellite altitude}\label{Dif_hs_1} 

In this subsection, all the orbital perturbation effects considered are evaluated for a given satellite altitude.

\begin{figure}
\centering
\includegraphics[width=12cm,height=8cm]{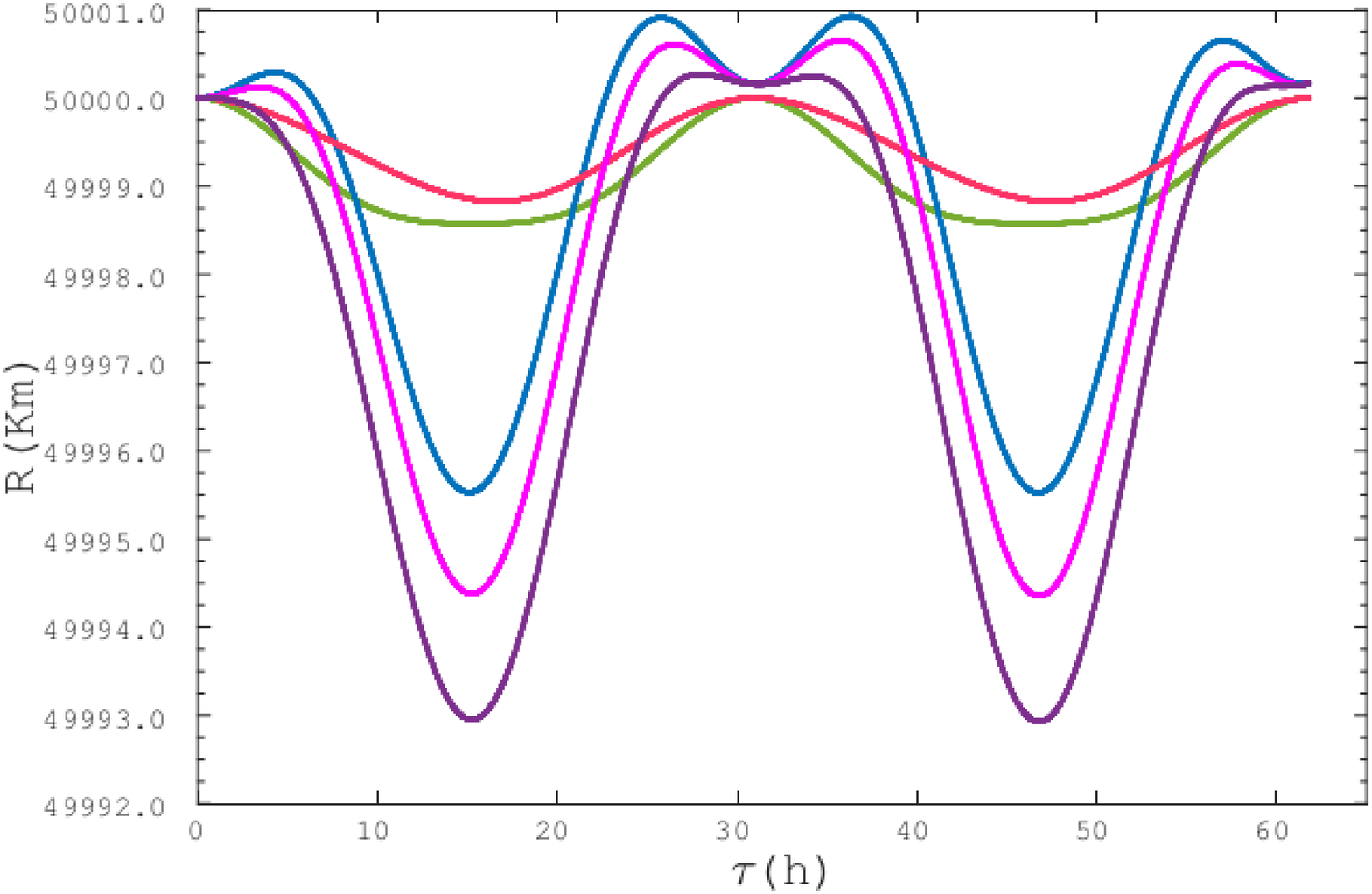}
\caption{Radial distance of a satellite versus the proper time (for two orbital periods) at $5 \times 10^{4} \ Km$ from the geocenter. 
The orbital perturbation effects that are considered in the metric are Earth oblateness ($green$), Moon gravitational potential ($blue$), Sun gravitational potential ($red$), 
Moon plus Sun gravitational potential ($magenta$) and, Earth oblateness plus Moon plus Sun gravitational potential ($violet$).}
\label{fig_all_efs_ephemerides_50000km}
\end{figure}

Figs.~\ref{fig_all_efs_ephemerides_50000km} and \ref{fig_100000_150000_200000}
show the radial distance of a satellite versus the proper time (for two orbital periods), 
from the geocenter, at satellite orbital radius $5 \times 10^{4} \ Km$, $10^{5} \ Km$ and $1.5 \times 10^{5} \ Km$. 
The perturbing orbit effects that are considered in the metric are Earth oblateness, Moon gravitational potential, 
Sun gravitational potential, Moon plus Sun gravitational potential and, Earth oblateness plus Moon plus Sun gravitational potential.     

From Fig.~\ref{fig_all_efs_ephemerides_50000km}, we can conclude that when the satellite is orbiting at $5 \times 10^{4} \ Km$ 
the perturbing effects considered such as Earth oblateness, 
Moon gravitational potential or Sun gravitational potential are of the same order of magnitude, but 
the Moon's effect is the strongest one 
(see the amplitude between maximum and minimum distance to the geocenter 
for the cases when the Moon gravitational potential is included). 
Note that the Earth oblateness effect is bigger than Sun and Moon effect at the height of Galileo satellites, 
but this is not the case at $5 \times 10^{4} \ Km$ of orbital radius. All these results are compatible 
with the figure which gives the order of magnitude of various perturbations of a satellite orbit from the book \cite{mon05}.

From Fig.~\ref{fig_100000_150000_200000}, we can conclude that as the orbital radius of the satellite increases, 
the Earth oblateness effect decreases and the Moon and Sun effects increase. When the satellite is approaching the Moon, the Moon effect is bigger than the Sun one.

\FloatBarrier

\subsection{Considering one orbital perturbation effect varying the satellite altitude}\label{Dif_hs_2} 

In this subsection, different orbital perturbation effects (Moon plus Sun gravitational field and Earth quadrupole) are 
evaluated for different satellite altitudes.

Fig.~\ref{fESM_dif_h} 
shows the radial distance, divided by its given orbital radius, for a satellite, versus the proper time (for two orbital periods),  
at different orbital radius from the geocenter. 
Notice that now a relative distance variation is shown so as to describe the distance variation 
from the different satellite positions to 
the geocenter. The orbital radius is the nominal one, a circumference
described by Schwarzschild metric. 
The orbit perturbations that are taken into account in the metric are Moon plus Sun gravitational potential. 
The Earth oblateness effect is not shown because it decreases when the distance from the geocenter increases. This effect is much smaller than Moon and Sun 
gravitational potential effects for orbital radius greater than about $5 \times 10^{4} \ Km$.

From 
Fig.~\ref{fESM_dif_h}, 
we can conclude that the perturbations of a satellite orbit given 
by Moon plus Sun gravitational potentials increase as the satellite orbital radius increases. 

Fig.~\ref{fig_ESM_ephemerides_dif_h_J2} shows the radial distance, divided by its given orbital radius, for a satellite, 
as a function of the proper time (for two orbital periods),
at different orbital radius from the geocenter. 
The orbital perturbation effect that
is considered in the metric is the Earth oblateness.

Fig.~\ref{fig_ESM_ephemerides_dif_h_J2} clearly shows that when the orbital radius of the satellite is increasing the Earth oblateness effect is decreasing. 
This effect is significant for satellites with lower orbital radius.

\FloatBarrier
\section{Relativistic positioning and user location}\label{RPS}

In this section, the U-errors are computed
taking into account the difference in location when
considering Schwarzschild metric and the metric
we have introduced. 
In this way we can appreciate the difference
in the positioning with the
two metrics. The description of the
satellite world lines taking 
into account the Moon, the Sun and 
the Earth quadrupole is more precise 
and is closer to the positioning it
could be expected from the Galileo Satellite Constellation.  

As stated before, the computations presented
here are numerically calculated 
implementing known analytical results by \cite{col10a} and \cite{col10b} concerning RPS. 

Our computations are performed in a similar way as in \cite{neu14}. Let us then describe the
main features of such computations. 
Firstly, in subsection \ref{RPS1}, two codes ($XTcode/TXcode$ ) are presented, 
as well as their implementation to calculate
the positioning errors. 
Secondly, in subsection \ref{RPS2}, the procedure to estimate the positioning errors is described. 
Thirdly, in subsection \ref{RPS3}, the HEALPIx representation and the initial users distribution are explained. 
This kind of representation is used to display our numerical results.
Fourthly, in subsection \ref{RPS4}, the analysis of the numerical results is done.
Finally, in subsection \ref{RPS5}, the spatial configuration of user-satellites that is associated to maximum positioning errors is described.

\subsection{$XTcode/TXcode$}\label{RPS1}



The positioning of a user in space-time is done by two calculations: 1) satellite geodesics and 2) photon geodesics.

First the satellite world lines have to be calculated. 
Once this is done the user location is obtained in a certain space-time. 
It is necessary to solve the null geodesics followed by the photons in such space-time. 
The 
user inertial coordinates, $x^{\alpha}$, are defined, and the emission ones, $\tau^{A}$ 
(the proper times that the user receive at the same time from the four satellites, 
remember that $A$ is the
number which describe each of the four satellites). In Minskowski space-time 
those coordinates must satisfy the following algebraic equations:
\begin{equation}
\eta_{\alpha \beta} [x^{\alpha} - x^{\alpha}_{A}(\tau^{A}) ]
[ x^{\beta} - x^{\beta}_{A}(\tau^{A}) ] = 0   \ .
\label{cap1-7}
\end{equation}
where $\eta_{\alpha \beta}$ is the Minkowski diagonal matrix with $\eta_{00}=-1$ and $\eta_{11}=\eta_{22}=\eta_{33}=1$, 
and the points of the satellite world
lines have inertial coordinates $x^{\beta}_{A}(\tau^{A})$,
which must be
well known functions of the proper times $\tau^{A}$. According
to Eqs.~(\ref{cap1-7}), photons follow null geodesics from satellite
emission to user reception. These algebraic equations
may be solved by the knowledge of the satellite world lines. Also, a numerical Newton-Raphson method (\cite{pre99}) is used.
(See \cite{neu12} and \cite{neu14} for a more detailed description). 
 
We can proceed in two senses.
Solve Eqs.~(\ref{cap1-7}) for known $x^{\alpha}$ and determine  $\tau^{A}$
or inversely. In the first case, 
Eqs.~(\ref{cap1-7}) may be numerically solved for the unknowns $\tau^{A}$
 by assuming that the position coordinates $x^{\alpha}$ are
known. Thus, the emission coordinates are obtained
from the inertial ones. However, the same equations
may be solved to get the unknowns $x^{\alpha}$ for known emission coordinates $\tau^{A}$. This second case gives the 
inertial coordinates in terms of the emission ones (positioning); nevertheless, this second numerical solution
of Eqs.~(\ref{cap1-7}) is not necessary since there is an analytical
formula obtained in \cite{col10a}, which gives $x^{\alpha}$ in terms of $\tau^{A}$ 
for photons moving in Minkowski space-time, and arbitrary satellite
world lines.

In practice, a numerical code with multiple precision was
designed 
to calculate the emission coordinates $\tau^{A}$
(unknowns) from the inertial ones (data) by solving
Eqs.~(\ref{cap1-7}). It is hereafter referred to as the {\it XT-code}.
This code, based on the Newton-Raphson numerical
method, requires the satellite world line equations; that
is to say, there must be a subroutine which calculates
the inertial coordinates of every satellite $x^{\alpha}_{A}(\tau^{A})$ for any
value of $\tau^{A}$ (see Appendices A and B).
Moreover, we built up a numerical code, based
on the analytical formula obtained in \cite{col10a}, which, for given emission coordinates $\tau^{A}$,
allows us the calculation of the user inertial coordinates $x^{\alpha}$, positioning. This code is hereafter referred to as the
{\it TX-code}. 


\FloatBarrier
\subsection{Effects of the metric on the positioning}\label{RPS2}

In this subsection, the procedure to calculate the positioning errors is described.

Let us first suppose that the satellites move without uncertainties (without any perturbation effect). Their trajectories are circumferences as
it corresponds to the Schwarzschild space-time (see Appendix B). These
trajectories are circular orbits in the case
of a spherically symmetric non rotating Earth, in the
absence of external actions (nominal trajectories).

In practice, any realistic satellite world line deviates with respect to the
nominal ones. If the nominal world lines are parametrized by means of their proper times,
the equations of these world lines (see Appendix B) may be written as follows: $x^{\alpha}_{A}(\tau^{A})$ being $x^{\alpha}_{A}$ ($A=1...4$) the coordinates
of a given satellite $A$, which is a function of its proper time $\tau^{A}$.
In the present paper, the realistic perturbed
world lines are the timelike geodesics in the space-time calculated in the previous sections (see Appendix C). The satellites are orbiting
at the height of the Galileo Constellation and, in the present work, only the following
major effects are considered in the metric: the Earth oblateness and the gravitational effects of the Earth, the Moon and the Sun.

In the absence of deviations with respect to the nominal lines, our
{\it XT-code} gives the emission coordinates $\tau^{A}$ corresponding to any set of inertial coordinates $x^{\alpha}$ and, then, from
the resulting emission coordinates, the {\it TX-code} allows us to recover the initial inertial ones. The number of digits
recovered measures the accuracy of our {\it XT} and {\it TX codes}. Since multiple precision is used, this accuracy is
excellent.

Let us now take the above emission coordinates $\tau^{A}$,
which are not to be varied since they are broadcasted by
the satellites and received by the user without ambiguity. For these coordinates and the perturbed satellite world
lines in the space-time (including the perturbations effects: 
Moon and Sun gravitational field and the Earth oblateness) calculated in the previous sections, 
the {\it TX-code} gives new inertial coordinates 
$x^{\alpha}+\Delta (x^{\alpha})$. Coordinates $x^{\alpha}+\Delta (x^{\alpha})$ are to be
compared with the inertial coordinates $x^{\alpha}$ initially assumed. 
The quantity 
\begin{equation}
\Delta_{d}=[\Delta^{2}(x^{1})+\Delta^{2}(x^{2})+\Delta^{2}(x^{3})]^{1/2}
\label{delta_d}
\end{equation}
is a good estimator of the positioning errors produced
by the perturbations of the satellite motions. Those errors are called U-errors. 
It is worthwhile to emphasize that user positions $x^{\alpha}$ and $x^{\alpha}+\Delta (x^{\alpha})$
correspond to the same emission coordinates, which are received from the satellites, but
to different world lines. The nominal world lines lead to
$x^{\alpha}$ and the perturbed ones give $x^{\alpha}+\Delta (x^{\alpha})$.
We may then say that the user position is $x^{\alpha}$ with an error whose amplitude is given by the estimator $\Delta_{d}$. 
See \cite{neu14} for a more detailed explanation of the
definition of these U-errors. The improvement presented here is the use of a most accurate description  
of satellite perturbations using a metric which better accounts of a more accurate trajectory of the satellites. 

\FloatBarrier
\subsection{HEALPIx representations and initial users distribution}\label{RPS3}

In a similar way as it was performed in \cite{neu14}, we are going to present the results of the 
computation of the $\Delta_{d}$, the U-errors, of different users equidistributed in
sphere surfaces with different radius centred in the geocenter. 

In order to represent such values an appropriate method is needed. This method allows us
to represent some quantities in 3D, $t \ =$ constant, space-time sections. An appropriate pixelization 
and colour bars are needed. 

The HEALPIx (Hierarchical Equal Area Isolatitude Pixelisation of the Sphe\-re) package
(\cite{gor99}) is a very useful tool for this representation 
and it is used here. This package was initially designed to represent the Cosmic Microwave Background temperature distribution in the sky.
As it writes the values of a scalar quantity in a sphere surface, it is also very convenient for our kind
of pictures. This method displays any quantity as a function of the direction (pixel). The sphere surface is divided in
$12 \times N^2_{side}$ pixels and the free parameter $N_{side}$ takes even natural numbers. In order to 
better compare with the pixelization considered in \cite{neu14} $N_{side} = 16$ is taken. 
Initial users are fixed on a surface of a sphere of radius $R$ centred in the geocenter. 
One user per each HEALPIx direction. 3072 initial users (pixels) equally distributed are considered. 
The angular area of any one of them is $\approx 13.43$ square degrees. Such angular area is close to sixty four times
the mean angular area of the full moon. However their shape in the mollweide representation used here is not the same for all pixels.
They are more elongated in the polar zone. 
For each initial user, we carry out the procedure that we have commented 
for the computations.\\ 
The pixelized sphere is shown by using the mollweide projection, in which, the frontal hemisphere is projected on the central part of the figure, and the opposite 
hemisphere is represented in the lateral parts.

As stated in the latter paragraph, 
HEALPIx mollweide maps are taken to represent the scalar quantity required, in our case the U-errors, $\Delta_{d}$. 
According to the colour bar displayed, any pixel shows a colour which states the $\Delta_{d}$ value in the map
for the direction associated to such pixel. 
$\Delta_{d}$ values for different cases 
are shown in Figs.~\ref{maps_RPS1}, \ref{maps_RPS2} and \ref{maps_RPS3}, considering Galileo satellites 2, 5, 20 and 23, in a similar way 
as in \cite{neu14}.\\
HEALPIx mollweide maps of Figs.~\ref{maps_RPS1} and \ref{maps_RPS2} are $\Delta_{d}$ values, in Km, on spheres with Earth radius $6378=R_{\oplus}$ at different user times $t$.
Otherwise, Fig.~\ref{maps_RPS3} shows $\Delta_{d}$ values, in Km, on spheres with different radius at user time $t=19 \ h$. 
From top to bottom, the radius of the spheres in kilometres are $1{.}5 \times 10^{4}$, $3 \times 10^{4}$ and
$5 \times 10^{4}$. The first and the third values are two of the values taken in \cite{neu14}.
This fact allows us to compare the results
obtained with the two different procedures to compute the U-errors.

\FloatBarrier
\subsection{Analysis of results (Figs. \ref{maps_RPS1} and \ref{maps_RPS2})}\label{RPS4}

In this subsection, the analysis of the numerical results for different initial users distributions is done. 
As stated at the end of the previous section, sphere surfaces at the Earth radius $6378=R_{\oplus}$ are taken. 
We want to know which is the U-error, $\Delta_d$, produced on the positioning taking into account the
deviation caused on nominal orbits (Schwarzschild ones) when the effects of Moon, Sun and Earth quadrupole 
are considered in the metric. In \cite{neu14} it was pointed out that for $6378=R_{\oplus}$ sphere radius 
the positioning errors where of the same order as the satellite deviations from the nominal orbits. 
In the cited paper, the deviation was statistically generated. Now more realistic perturbations are considered. 
Satellites with different relative positions are considered at different times.

The results commented in the last paragraph are obtained 
from HEALPIx maps shown in Figs.~\ref{maps_RPS1} and \ref{maps_RPS2}. Let us make a list of our main conclusions:

\begin{itemize}
\item The greatest $\Delta_{d}$ values (red pixels) correspond to having the maximum radial distance deviation of the satellite 
for the case of the four chosen satellites (see Fig. \ref{fig_ESMJ2_ephemerides}, violet line). When the satellite trajectory deviation 
with respect to the Schwarzschild trajectory (considering circular orbit as nominal trajectory) increases, then 
the value $\Delta_{d}$ increases. For the positioning, the four satellites are needed. 
Therefore, this deviation for the four chosen satellites has influence on the positioning errors. 
So, the value $\Delta_{d}$ depends directly on the satellite-Earth-Moon-Sun spatial configuration, and this for the four satellites. 
\item The $\Delta_{d}$ values are of the same order of magnitude of the perturbation applied on the satellite orbit in most of the pixels. That is, 
about a few Kilometres is this order of magnitude, just the same as the satellite deviations when we take into account Earth oblateness plus Sun plus Moon gravitational field. 
These results are in agreement with the respective results obtained in \cite{neu14}.
\item Almost the same shape on the maps (but with different maximum and minimum values) is repeated when we consider the next satellite orbital period. 
This periodic effect was to be expected because the Sun-Moon-Earth-satellite spatial configuration practically does not change after one satellite orbital period ($14.2 \ h$). 
Therefore, Sun, Moon and Earth influence on the satellite orbit is almost the same than in the previous satellite orbital period. 
This periodic effect is shown in Figs. \ref{maps_RPS1} and \ref{maps_RPS2}, corresponding to the cases where the receiver time is 11h and 25h, respectively.  
\item The Earth oblateness (Earth quadrupole) effect is shown in Fig.~\ref{maps_RPS2} ($receiver$ $time=21 \ h$), as it can be seen in Eq.~(\ref{pertur_J2}), 
the sinusoidal shape is obtained. 
\end{itemize}

\FloatBarrier
\subsection{Jacobian and user-satellites spatial configuration (Fig.~\ref{maps_RPS3})}\label{RPS5}

In this subsection we present a case with a user-satellites 
spatial configuration that is associated to maximum positioning errors. The purpose
of it is to enhance the relationship of the U-errors with the Jacobian, $J$, values as
it was stated in the \cite{neu14} work. For our computation of the errors
based on a more realistic case, the relationship between the Jacobian values
near zero and the great values of the U-errors still stands, as it must be. 

For the sake of comparison the fig 2 top case of \cite{neu14} has been considered. 
Let us recall which is this Galileo satellites configuration: satellites 2, 5, 20
and 23 at user time $t=19 \ h$. 
Fig.~\ref{maps_RPS3} shows HEALPIx mollweide maps of $\Delta_{d}$ values, in Km, on spheres with different radius
for this satellite distribution. Two radius spheres in kilometres as in figures 5 and 6 of \cite{neu14}
have been considered:  $1{.}5 \times 10^{4}$ (top) and
$5 \times 10^{4}$ (bottom). Also $3 \times 10^{4}$ (middle) is taken here. 
Top subfigure can be compared with the top one of fig 6 in \cite{neu14}, the satellites
configuration and user time are the same ones. The difference
between both figures is the way of computing the $\Delta_d$ errors.
The distribution of those U-errors is very similar as it can be seen
comparing both subfigures. This is logical as the same satellites
configuration and user time have been taken. 
When we use our metric, lower errors are obtained.
The range in \cite{neu14} varies from $6.84$ to $24.0$ $km$, while
in the present paper the range of such values varies from  
$0.2$ to $10.1$ $km$. The lower values obtained in the 
present work should be due to our better approximation: taking 
a realistic metric better describes the U-errors in RPS. 

As in the mentioned paper, some cutoffs of $\Delta_d$ are considered. 
In Fig.~\ref{maps_RPS3},  grey coloured pixels are characterized by the condition $\Delta_{d} > 50 \ Km$ for the middle subfigure 
and $\Delta_{d} > 100 \ Km$ for the bottom subfigure. 
The same as in our previous work, as the radius of the sphere
increases the $\Delta_{d}$ values increase (more grey pixels). In the 
above cited paper it was explained the relation between the volume of the tetrahedron formed 
by the tips of the four user-satellites unit vectors, $V_{T}$, and the Jacobian, $J$.
Specifically, this volume is a sixth of the $|J|$ value: $V_T = |J|/6$ (see also \cite{lan99}). 

We have chosen a user whose $\Delta_{d}$ value is one of the greatest of Fig. \ref{maps_RPS3} ($R=3 \times 10^{4} \ Km$).
Fig.~\ref{celest_sphere_user} shows the locations of the four satellites in the celestial sphere of this user, whose user-satellites 
spatial configuration corresponds to a case of $J \simeq 0$ ($J=-6.7 \times 10^{-2}$). Let us recall that  
$J=|\partial \tau^{A}/\partial x^{\alpha}|$, of the transformation giving the emission coordinates in terms of the inertial ones. 
Note that the considered Galileo satellites are 2, 5, 20 and 23 at user time $t=19 \ h$ for a user on a surface of a 
sphere of radius $R=3 \times 10^{4} \ Km$. Remember that the volume $V_{T}$ is proportional to the Jacobian value. 
In Fig.~\ref{celest_sphere_user}, the satellites are represented by four black asteriks, the user by a blue cross (the origin of the Cartesian system) 
and the centres of the circles by red, blue, green and purple asterisks. Four different circles, 
that pass through three points, with each possible combination of three satellites (colour points) are also drawn. As it was explained
in \cite{neu14}, such satellite configuration corresponds to a tetrahedon near of null volume value and therefore to a $J$ close to zero.  
This fact accounts for such a great U-error value. 

In \cite{neu14} it was also studied the relation of the user distance to Earth and the $\Delta_d$ errors. Let us make
a comment here taken into account the figures presented in this work. 
From HEALPIx maps shown in Fig.\ref{maps_RPS3}, we also see that as the radius of the spherical surfaces 
increases the $\Delta_{d}$ values also increase (see how the minimum value of $\Delta_d$  increases
with $R$ in these maps). 
This is because if the user is very far from the satellites, 
they are all in a small solid angle and the tetrahedron volume $V_{T}$ is also expected to be small. 
Moreover, on these maps, there are regions with $J \simeq 0$ (grey pixels with high $\Delta_{d}$ value) 
when the height is greater than approx. $2 \times 10^{4} \ Km$.  
In \cite{neu14}, it was also shown that not only regions near $J$ null values give great $\Delta_d$ error values. 
There are also regions where $\Delta_d$ is small.
One of the
conclusions in the mentioned paper was that 
to solve this problem of big positioning errors at great distances, it should be interesting in the future, 
to choose the best combination of four satellites (from GPS and Galileo constellations), 
whose user-satellites spatial configuration gives smaller $\Delta_d$ errors. This fact seems
possible if the satellites configurations taken into account are changed accordingly.
Some proposals of how to determine such configurations can be seen in \cite{neu14} and
also apply to the results obtained in this paper. 
Another way to do that could be to locate the satellites in other orbits in the solar system, 
in a way that the user-satellites spatial configuration corresponds to $V_{T}$ values not close to zero.


\FloatBarrier

\section{Conclusions and perspectives}\label{conclusions}

The main purpose of this paper is to study the U-errors obtained as the difference in RPS by 
using Schwarzschild and a more accurate metric to describe the Galileo satellite world lines.
This represents an advance with respect to the research made by \cite{neu14}. In \cite{neu14}
the satellite perturbations were statistically computed. Here such
perturbations are computed differently by taking into account the effects of 
Moon, Sun and Earth quadrupole in the metric. These are the greatest effects perturbing satellite
world lines at the height of Galileo or GPS constellation.
In order to better understand the difference in the positioning, the U-errors,
a deep study of the change in the Galileo satellites trajectories has been first developed, 
see sections \ref{pert_ef} and \ref{Dif_hs}. This study takes into account the contribution on the satellite world lines of such three
terms by (i) taking them separately, (ii) combining two of them and (iii) considering all three
together.

A Runge-Kutta
algorithm is used to solve the geodesic equations. High accuracy is achieved
($10^{-18}$). Also multiple precision (40 significant digits for each number) is used. 
Adequate initial conditions have been found to solve the ODE. 
Precise satellite geodesics are required to be able to precisely compute 
the user location in space-time. Moreover, this precision is needed to incorporate
other small contributions to the satellite orbits perturbations.
In this work, only the Galileo Satellite Constellation is considered. As it is known, one of the greater
contributions at the height of Galileo satellites 
is the Earth oblateness (about a few kilometres). The Moon and Sun gravitational effects
are also important (about a few hundred Kilometres). Moreover, due to the greater relative Moon motion, 
after one orbital period, the satellite position is shifted when the Moon is considered.
Our results are in agreement with other results known in the literature. 
Therefore, our method is working properly because  
the order of magnitude computed in preceding works with the GNSS approximation is obtained (see for instance 
\cite{teu15}, Chapter 3, \cite{mon05} and \cite{Kyo18}). 
The RPS methods are more exact than GNSS classical procedures, even when the last ones 
incorporate relativity corrections. 
From the beginning, we numerically solve the satellite geodesic equations (ODE) in our given space-time.

Then we have simulated satellite world lines at different distances from Earth (see section \ref{Dif_hs}) and
studied the influence of the perturbations considered in our metric. 
The Earth oblateness orbit perturbing contribution is of the same order of magnitude than that of Moon and Sun gravitational effect at $5 \times 10^{4} Km$. 
As the satellite orbital height increases, the Earth oblateness effect decreases and Moon and Sun effects increase. 
At $10^{5} Km$, the Earth oblateness effect is smaller than the Moon and the Sun contributions. 
At these distances, the satellite position is also shifted when the Moon presence is considered, as it happens at the Galileo satellites distances.
This is an improvement with respect to our previous works because now it is possible to appreciate the separate 
contribution of each perturbation in the RPS.

After studying the satellite trajectories, we carry out RPS with our metric. 
A similar procedure as in \cite{neu14} has been performed. 
But now, a metric taking into account the greater 
physical perturbations at Galileo Satellite Constellation is considered
to compute $\Delta_d$. Notice that the same formula, \cite{col10a}, has
been used to compute the proper times that the user receives from the 
satellites. Also HEALPIx maps are considered to describe
the U-errors as in \cite{neu14}. 

As it was concluded in the cited paper, the positioning errors values, $\Delta_{d}$, 
are almost of the same order of magnitude as those of the perturbed satellite orbits (orbital perturbation effect).
Here the highest  $\Delta_{d}$ values (red pixels in subsection \ref{RPS4}) correspond to having the maximum radial distance deviations of the satellite for the case of the four chosen 
satellites in Fig. \ref{fig_ESMJ2_ephemerides} (violet line). 
So, the value $\Delta_{d}$ 
depends directly on the satellite-Earth-Moon-Sun relative spatial configuration and it does so for each of the four satellites.  
Almost the same HEALPIx maps are recovered after a Galileo satellite orbital period. 
This is because the relative spatial configuration among satellite-Moon-Sun-Earth does not nearly change after $14.2 \ h$ (periodic effect), 
as the Moon and Sun hardly move after a Galileo orbital period. Here the $\Delta_d$ values are smaller than the ones
obtained with the statistical procedure used in \cite{neu14}. This fact is due to a more realistic representation of 
the satellite orbital perturbations by the use of metrics instead of statistical deviations (see subsection \ref{RPS5}). 
Therefore a more accurate computation of the U-errors is performed and so a more precise calculation of
the user's positioning can be achieved.

As in \cite{neu14} when the radii of the spherical surfaces increase, the $\Delta_{d}$ values increase. 
This is because if the user is very far from the satellites, these are all in a small solid angle and the tetrahedron volume $V_{T}$ is expected to be small. 
Moreover, on these maps, there are regions with $J \simeq 0$ when the radius is greater than about $20000 Km$. 
One solution when $J \simeq 0$, is to choose another combination of four satellites from Galileo or GPS constellation, 
whose user-satellites configuration is associated to values of $J$ not close to zero. 
An alternative way to do that could be to locate the satellites in other orbits in the solar system, 
in a way that the user-satellites spatial configuration corresponds to $V_{T}$ values not close to zero.

We are currently working on an improvement in our numerical procedure.
The Newton-Raphson 
numerical method (see {\it XTcode} in section \ref{RPS1}) could be avoided, in such a way that analytical 
functions of the proper time are not used, since they imply the use of Schwarzschild world lines for the satellites. 
For example, to use the secant method with the satellite world lines (taking into account the Sun and Moon presence and the Earth quadrupole).
We think this will improve the numerical code and results. 

Once this is done, another interesting thing to do, and very useful, should be to create HEALPIx maps as 
in subsection \ref{RPS4} (see Figs.~\ref{maps_RPS1} and \ref{maps_RPS2}), but calculating the positioning error $\Delta_{d}$ on the geoid, 
instead of on the spherical surface of radius $R_{\oplus}$. It would allow us to calculate the accuracy of Galileo satellites, 
with our implementation, and compare with the data from Galileo Constellation, and other constellations. 
These results should also be interesting for geodesic treatment. 

There are other perturbations a part
from those considered in this paper that also contribute to the computation of the satellite world lines. The order of magnitude of 
such contributions depends on the satellite's height as it can be seen, for instance, in \cite{mon05} (see, in particular, fig 3.1 at page 55). 
Therefore, this variation of the trajectories of the four satellites considered should contribute to the change in the calculations of RPS.  We are studying
such contributions and will present the results obtained elsewhere.

Also the use of our method in space navigation is being planned. 
The Barycentric Celestial Reference System could be used as reference system to locate the emitters (four satellites) 
in the solar system, in such a way that the configurations of the user-satellites associated to $J \simeq 0$ are avoided. 
For example, in the vicinity of the Moon,
two emitters fixed on the Moon surface (North and South poles) and two emitters from Galileo satellites. 
In such a case, the $V_{T}$ is not close to zero, in most of the cases. The positioning of 
a spacecraft that navigates in the solar system could be determined 
considering emitters in other appropriate locations to be studied.

\section*{Acknowledgments}

We would like to acknowledge our great debt to Professor 
Diego Pascual S\'{a}ez Mil\'{a}n 
who was the pioneer 
of this research and worked in its theoretical approach. 
We worked together for a very long time in much of our common projects. 
He left us three years ago. There are no words to describe our 
gratitude, both for his scientific and his human teachings.
We also acknowledge Dr. J.A. Morales-Lladosa for all his help. Dr. Pac\^ome Delva 
should also be mentioned for the same reason. 
This work has been supported by the Spanish Ministerio de Ciencia, Innovaci\'on y Universidades and the Fondo Europeo de Desarrollo Regional, 
Projects PID2019-109753GB-C21 and PID2019-109753GB-C22, the Generalitat Valenciana Project AICO/2020/125 
and the Universitat de Val\`encia Special Action Project  UV-INVAE19-1197312.






\bibliographystyle{agsm}



\appendix

\section{\bf Effects on the Galileo satellite world lines} \label{justification}   

So as to better understand the perturbations on the
satellite world lines, at the beginning of our work
we made some computations of the order
of magnitude of the 
effects of the Earth, Sun and Moon on Galileo satellites. 
These are the major perturbations on the satellite orbits at the satellite height (see \cite{Kyo18}). 
For this reason it is interesting to  
study the magnitude of their gravitational field and potential at approx. 
$3 \times 10^{4} \ Km$ from the geocentre (the radius of a Galileo orbit). Also  
the relation between them (Earth-Sun and Earth-Moon system) is relevant. 
In a near future the order of magnitude of other smaller 
effects will be studied. 
Notice that here we do not present the metric itself, we only compare the relation between the order of magnitude of Newtonian effects on satellite orbits, for the sake of 
better understanding the impact of the effects considered here. 

We assume that the geocentric position vector for Sun, Moon and satellite is
$\mathbf{x}_{\odot}=(-1, 0, 0) \ AU$, $\mathbf{x}_{\leftmoon}=(-384402.0, 0, 0) \ Km$ 
and $\mathbf{x}=(29655.3, 0, 0) \ Km$, respectively. In all the paper
$\odot,\oplus, \leftmoon$ stands for Sun, Earth and Moon respectively.\\
As it has been said the origin of the reference system is the geocentre (GRS). 
Afterwards, when we consider the Earth movement and acceleration, the GRS should be co-moving (the origin of the GRS is moving with the Earth centre).

The quantities computed here are the order of magnitude of
Newtonian potentials and accelerations on the Galileo satellites positions. 

Two cases are computed: 
\begin{enumerate}[({\bf a})]
\item {\bf Stationary Earth. 
\item Earth with accelerated motion and the reference system co-moving with the Earth.}
\end{enumerate}

For the (a) case, we calculate:
\begin{itemize}
\item The Earth potential divided by the Sun potential: $\frac{\phi_{\oplus}}{\phi_{\odot}}=0.0150$. The result is that the Sun potential is 67 times greater than the Earth's potential. 
\item The intensity of the Earth gravitational field divided by the intensity of the Sun's gravitational field is: $\frac{a_{\oplus}}{a_{\odot}}=74.6525$. 
Then, the force of the Sun on the satellite is 75 times smaller than the force of the Earth on it.

Although the Sun potential has a great effect, its variations, $a_{\odot}$, corresponding to the forces which represent the gravitational field,  
can be neglected in $a_{\oplus}$ and the quantity $ \phi_{\odot}$ 
could be taken as an additive constant of the $ \phi_{\oplus}$. 
 
\item Moon potential divided by the Earth's potential: $\frac{\phi_{\leftmoon}}{\phi_{\oplus}}=0.0010$. This  effect ($\phi_{\leftmoon}$) is either negligible or 
could be taken as a perturbation.
\item The intensity of the Moon gravitational field divided by the intensity of the Earth's: $\frac{a_{\leftmoon}}{a_{\oplus}}=8.8139 \times 10^{-5}$. 
This quantity ($a_{\leftmoon}$) could be neglected (as we do) or considered as a small perturbation.
\end{itemize}

The Newtonian concept of force cannot be applied to GR. 
The Christoffel symbols take the "role" of gravitational forces. 
In the set of ODE, Eqs.~(\ref{eq_41})-(\ref{eq_44}), as it can be seen in Appendix~\ref{EDOs}, 
the Sun, Earth and Moon "gravitational potentials" are placed in the denominator. 
And the term $1+\partial \phi_{\oplus}+\partial \phi_{\odot}+\partial \phi_{\leftmoon}$ is in the numerator.
We next present a summary of the results obtained for case (a):

Results of case (a): 
\begin{itemize}
\item $a_{\oplus} >> a_{\odot}$
\item $\phi_{\oplus}<<\phi_{\odot}$
\item $a_{\oplus} >> a_{\leftmoon}$
\item $\phi_{\oplus}>>\phi_{\leftmoon}$
\end{itemize}
However, the Earth is accelerated [(b) case] under the action of the celestial body (Sun or Moon) and 
such action will be included now.
 
In the book \cite{teu15} (Chapter 3), the authors show how to compute such perturbations. 
The following equation is considered: 
\begin{equation}
\mathbf{a}_{B}=G M_{B} \left(\frac{\mathbf{x}_{B}-\mathbf{x}}{|\mathbf{x}_{B}-\mathbf{x}|^{3}}-\frac{\mathbf{x}_{B}}{|\mathbf{x}_{B}|^{3}}  \right) \equiv \mathbf{a}_{B1}+\mathbf{a}_{B2}.
\label{eqbook_GNSS2}
\end{equation}
$M_{B}$ is the mass of the celestial body $B$ (Sun or Moon), $\mathbf{x}_{B}$ its geocentric position and $\mathbf{x}$
is the geocentric position vector of the satellite.

The term $\mathbf{a}_{B1}$ of Eq.~(\ref{eqbook_GNSS2}) is:  
\begin{equation}
\mathbf{a}_{B1}=G M_{B} \left(\frac{\mathbf{x}_{B}-\mathbf{x}}{|\mathbf{x}_{B}-\mathbf{x}|^{3}} \right) 
\label{term1}
\end{equation}
 This term $\mathbf{a}_{B1}$ stands for
the perturbing acceleration of $B$ exerted on the satellite.
And the term $\mathbf{a}_{B2}$ of Eq.~(\ref{eqbook_GNSS2}) is: 
\begin{equation}
\mathbf{a}_{B2}=-G M_{B} \left(\frac{\mathbf{x}_{B}}{|\mathbf{x}_{B}|^{3}}  \right) 
\label{term2}
\end{equation}
This term $\mathbf{a}_{B2}$ is the perturbing acceleration (on the satellite) 
caused by the action of $B$ on
the Earth. This is the inertial contribution (notice the minus sign) caused by the accelerated GRS.\\ Accelerations ($\mathbf{a}_{B1}$ and $\mathbf{a}_{B2}$) on the orbit of a Galileo satellite due to the Sun and the Moon are computed and their values can be seen in Table \ref{table_perturb}. 
Let us recall that the geocentric position vectors for Sun, Moon and satellite are 
$\mathbf{x}_{\odot}=(-1, 0, 0) \ AU$, $\mathbf{x}_{\leftmoon}=(-384402.0, 0, 0) \ Km$ and $\mathbf{x}=(29655.3, 0, 0) \ Km$, respectively (satellite-Moon-Sun aligned in the x axis). 
Table \ref{table_perturb} shows
the first Cartesian component of $\mathbf{a}_{B}$. 
The numerical results of (b) appear in the fourth column of Table \ref{table_perturb}. 

Nevertheless, when we consider a stationary Earth [(a) case], the term $\mathbf{a}_{B2}$ is zero. The results in case (a) appear in the second column of Table \ref{table_perturb}.
\begin{table}
 \centering  
\begin{tabular}{|c|r|r|r|}                                                                           
\hline                                                                                             
$Perturbation$ & ${a}_{B1} \ (m/s^{2})$ & ${a}_{B2} \ (m/s^{2})$ & ${a}_{B} \ (m/s^{2})$ \\                                                         
\hline                                                                                             
\hline                                                                                             
 $\leftmoon$ & $-2.860 \times 10^{-5}$ & $3.318 \times 10^{-5}$ & $4.583 \times 10^{-6}$ \\
 $\odot$ & $-5.928 \times 10^{-3}$ & $5.930 \times 10^{-3}$ & $2.350 \times 10^{-6}$ \\
 $J_{2}$ &                       &                      & $-6.793 \times 10^{-5}$ \\
\hline                   
\end{tabular}                                                                                      
\caption{Perturbing acceleration, for the first Cartesian component, on the Galileo satellite orbits for the effects: gravitational field from Sun and Moon, and Earth quadrupole.}
\label{table_perturb}                                                                                       
\end{table}       

We can summarize the results obtained and presented in Table~\ref{table_perturb}
in this way: i) the perturbing acceleration from the Sun is slightly smaller than that of the Moon. ii) Meanwhile, the terms $\mathbf{a}_{B1}$ and $\mathbf{a}_{B2}$ from the 
Sun are 100 times greater than those from the Moon. 
iii) As a first conclusion, we observe that the intensity of the Sun gravitational field is smaller than such from the Moon at the GRS that considers the Earth accelerated 
by the gravitational force of the Sun and the Moon. 

On the one hand, a stationary Earth (in a GRS), gives an
acceleration from the Sun 100 times greater than that from the Moon.
On the other hand, we have obtained an acceleration from the Earth central force (for the first Cartesian component) of $-0.453 (m/s^{2})$, 
considering $-G M_{\oplus} \left(\frac{\mathbf{x}}{|\mathbf{x}|^{3}}  \right)$ (acceleration of the Earth acting on
the satellite), in the motion equation of the satellite. 
Therefore the acceleration from the central force is $10^{5}$ times greater than the perturbing acceleration caused by Sun and Moon together.

The effect of the oblateness of the Earth is now studied. Its Cartesian components $(a_{1}^{J_{2}}, a_{2}^{J_{2}}, a_{3}^{J_{2}})$ with regard to its acceleration, 
\citep{see03, shar19}, are given by
\begin{eqnarray}
a_{i}^{J_{2}}=-3 G \left ( \frac{J_{2} M_{\oplus} \left (x^{i}-x_{\oplus}^{i} \right ) R_{\oplus}^{2} }{r_{\oplus}^{7}}\right ) \times \nonumber \\
\left[ A(i) \sum_{j=1}^{3} \left (x^{j}-x_{\oplus}^{j} \right )^{2}-5\left (x^{3}-x_{\oplus}^{3} \right )^{2} \right]
\label{pertur_acceler_J2}
\end{eqnarray}
where $A(i)$ is equal to 1 for $i=1,2$ and $A(i)$ is equal to 3 for $i=3$.
 

The perturbing acceleration from the Earth oblateness (for the first Cartesian component), $a_{1}^{J_{2}}$, 
has been computed and its value is $-6.793 \times 10^{-5}(m/s^{2})$ for $\theta=0$, as it is shown in the third row of Table \ref{table_perturb}. 
Let us recall that the satellite position vector is $\mathbf{x}=(29655.3, 0, 0) \ Km$.\\

As it can be seen in the fourth column of Table \ref{table_perturb}, the Earth oblateness produces an acceleration on the orbits of the 
Galileo satellite one order of magnitude greater, in absolute value ($6.793 \times 10^{-5}$) , than those produced by the Moon potential
($4.583 \times 10^{-6}$) or the Sun potential ($2.350 \times 10^{-6}$). 

In this Appendix, we have stated some important relations. They have helped us to better interpret 
the results obtained in our work.

\FloatBarrier
\section{\bf A short description of the 
circular satellite world lines } \label{EDOs}

As it is well known, in the unperturbed Minkowskian space-time,
the spatial location of a Galileo satellite $A$,
which moves along a circumference, requires 
three angles. Two of them, $\Theta $ and $\psi $, characterize 
one of the three orbital planes. These angles are 
constant. The third angle, $\alpha_{A} $, localizes the satellite
on its trajectory. 
It depends on time. All this was taken into account in \cite{neu12,neu14} 
to find  
the world line equations of the satellite $A$
to first order in the small dimensionless parameter
$GM_{\oplus}/R$, whose maximum value is $GM_{\oplus}/R_{\oplus} \simeq
6.94 \times 10^{-10} $. These equations are as follows:
\begin{eqnarray}
x^{1}_{A} &=& R \, [\cos \alpha_{A}(\tau) \cos \psi + \sin \alpha_{A}(\tau) 
\sin \psi \cos \Theta] \nonumber \\ 
x^{2}_{A} &=& -R \, [\cos \alpha_{A}(\tau) \sin \psi - 
\sin \alpha_{A}(\tau) \cos \psi \cos \Theta] \nonumber \\
x^{3}_{A} &=& -R \sin \alpha_{A}(\tau) \sin \Theta \nonumber \\
x^{4}_{A} &=&  \gamma \tau   \ ,
\label{satmot1}
\end{eqnarray} 
where the factor $\gamma $ and the angle $\alpha_{A}$ are given by the relations \citep{ash03,pas07}
\begin{equation}
\gamma = \frac {dt}{d\tau} = \Big( 1 - \frac {3GM_{\oplus}}{R} \Big)^{-1/2} \ 
\label{ttau} 
\end{equation}  
and 
\begin{equation}
\alpha_{A}(\tau) = \alpha_{A0} - \Omega \gamma \tau  \ ,
\label{satmot2} 
\end{equation}   
respectively. The last equation involves the 
satellite angular velocity
$\Omega = (GM_{\oplus}/R^{3})^{1/2} $, and the angle    
$\alpha_{A0}$ fixing the position of satellite $A$ at $\tau = x^{4} = 0$
(GNSS initial operation time). 
The chosen nominal world lines satisfy
Eqs.~(\ref{satmot1})--(\ref{satmot2}).

\FloatBarrier
\section{\bf Geodesic Equations of the Satellite world lines} \label{EDOs}
Now we present a description of the motion equations of a satellite in
the GR space-time considered in this work. 
The metrics that describe the contributions of Earth, Sun and Moon in GR are presented.
The solution of the time-like geodesics, equations of motion, are also described.
An overview of the numerical algorithm used in our solution is also explained. 

The International Astronomical Union IAU 2000 (see \cite{IAU2000, sof03}) recommends a resolution for the metric
tensor in the Geocenter Celestial Reference System GCRS up to order $o(1/c^{2})$ (neglecting terms smaller than $o(1/c^{2})$, for the $A$ $o(1/c^{3})$ body vector potential), such as:
\begin{eqnarray}
g_{00}=-\left[1-2 \left( w_{0}(t,\mathbf{x})+w_{L}(t,\mathbf{x})\right) \right],  \\ 
g_{0i}=0, \\ 
g_{ij}=\delta_{ij}\left[1+2 \left( w_{0}(t,\mathbf{x})+w_{L}(t,\mathbf{x})\right) \right]
\label{eq_2}
\end{eqnarray}
where $(t \equiv geocentric \:\:\:\: coordinate \:\:\:\: time, \mathbf{x})$ are the pseudo-Cartesian isotropic GCRS coordinates, $w_{0}=G \sum_{A} M_{A}/r_{A}$ 
summation over all solar system bodies, $ \mathbf{r}_{A}=\mathbf{x}-\mathbf{x}_{A}$, $\mathbf{x}_{A}$ is the position vector of the centre of mass of $A$ body, 
$r_{A}=|\mathbf{r}_{A}|$, and where $w_{L}$ has the expansion in terms of multipole moments that requires each body. 

In this paper only the masses $M_{A}$ of three bodies are considered: Sun, Earth and Moon ($A=\odot,\oplus, \leftmoon$). 
Although there are neglected terms with order greater than $\frac{GM_{A}}{r_{A}}$, they will be considered in the future.  
$w_{L}$ has been considered, being $2 w_{L}=\phi^{J_{2}}$ and $\phi^{J_{2}}$ the Earth quadrupole potential.
The quantity $w_{0}$ as $w_{0}=G \left( \frac{M_{\oplus}}{r_{\oplus}}+\frac{M_{\odot}}{r_{\odot}}+\frac{M_{\leftmoon}}{r_{\leftmoon}} \right)$ is also taken.
$\phi$ is defined as $2 \left( w_{0}(t,\mathbf{x})+w_{L}(t,\mathbf{x})\right)$. 
 

A numerical integration of timelike geodesic equations is performed to compute 
the satellite trajectories $x^{\alpha}(\tau)$ in the GCRS,
\begin{equation}
\frac{du^{\alpha}}{d\tau}=-\Gamma^{\alpha}_{\mu\nu}u^{\mu}u^{\nu}
\label{eq_3}
\end{equation}
where $u^{\nu}$ is the four-velocity ($u^{\nu}=\frac{dx^{\nu}}{d\tau}$) and $\tau$ the proper time.

The covariant and contravariant components of the metric in pseudo-Cartesian isotropic GCRS coordinates have this form:
\begin{eqnarray}
g_{\mu \nu}=\eta_{\mu \nu}+\phi \delta_{\mu \nu} \label{m_GCRS_cov} \\
g^{\mu \nu}=\eta^{\mu \nu}-\phi \delta^{\mu \nu}
\label{m_GCRS_contra}
\end{eqnarray}
and the Christoffel symbols are calculated from this definition:
\begin{equation}
\Gamma^{\alpha}_{\mu\nu}=\frac{1}{2} g^{\alpha \delta} \left( g_{\delta \nu,\mu} + g_{\mu \delta,\nu} -g_{\mu \nu,\delta} \right)
\label{eq_sym_christo}
\end{equation}
with:
\begin{equation}
g_{\mu \nu,\delta}=\phi_{,\delta} \delta_{\mu \nu}.
\label{der}
\end{equation}
being $\zeta_{,\alpha} \equiv \frac{\partial \zeta}{\partial x^{\alpha}}$, 
and $\zeta$ a function of $x^{\alpha}$.


The potential $\phi=\phi(x^{i})$ is considered as stationary. The non-null 
Christoffel symbols are  then obtained:
\begin{eqnarray}
\Gamma^{0}_{0i}=-\frac{1}{2}\frac{1}{1-\phi}\phi_{,i} 
\label{sc11} \\
\Gamma^{k}_{00}=-\frac{1}{2}\frac{1}{1+\phi}\phi_{,k} \\
\Gamma^{k}_{ij}=\frac{1}{2}\frac{1}{1+\phi}(\delta_{ik}\phi_{,j}+\delta_{kj}\phi_{,i}-\delta_{ij}\phi_{,k})
\label{sc}
\end{eqnarray}

The Christoffel symbols $\Gamma^{k}_{ij}$ are as follows:
\begin{eqnarray}
\Gamma^{i}_{ii}=\frac{1}{2}\frac{1}{1+\phi}\phi_{,i} \\
\Gamma^{i}_{jj}=-\frac{1}{2}\frac{1}{1+\phi}\phi_{,i} \;\;\;\;\;\; i \neq j \\
\Gamma^{i}_{ij}=\frac{1}{2}\frac{1}{1+\phi}\phi_{,j}  \;\;\;\;\;\;  i \neq j
\label{sc1}
\end{eqnarray}
(no summation over repeated indices).

The following expressions are the geodesic equations for these Christoffel symbols:
\begin{eqnarray} 
\frac{du^{0}}{d\tau}=\frac{1}{1-\phi}\phi_{,i}u^{0}u^{i} \label{eq_41} \\
\frac{dx^{0}}{d\tau}=u^{0} \label{eq_42} \\
\frac{du^{k}}{d\tau}=\frac{1}{1+\phi}\left[\frac{1}{2}\phi_{,k}(u^{0}u^{0}+u^{i}u^{i})-u^{k}(\phi_{,i}u^{i})\right] \label{eq_43}  \\ 
\frac{dx^{k}}{d\tau}=u^{k} \label{eq_44}  
\end{eqnarray}

The system has the following constraint:
\begin{equation}
g(u,u)=-1
\label{eq_5}
\end{equation}
This constraint lets us verify the proper functioning of the code.
In each step of the numerical integration the code checks that Eq.~(\ref{eq_5}) holds. 
A Runge-Kutta method is applied to solve such equations. 


In order to completely describe the motion of the system 
considered here, the motion equations of the celestial bodies taken into account should also be added to the system
of Eqs.~(\ref{eq_41})-(\ref{eq_44}). The Newtonian equations are sufficient. 
The Newtonian Earth motion equations in the GCRS, for the system Earth-Moon-Sun, are given by: 
\begin{eqnarray}
\frac{dx^{k}_{\oplus}}{d\tau}=v^{k}_{\oplus} u^{0}  \label{eq_61EarthMoonSun} \\
\frac{dv^{k}_{\oplus}}{d\tau}=-G \frac{M_{\odot} (x^{k}_{\oplus}-x^{k}_{\odot})}{|\mathbf{x_{\oplus}}-\mathbf{x_{\odot}}|^{3}} u^{0}
-G \frac{M_{\leftmoon} (x^{k}_{\oplus}-x^{k}_{\leftmoon})}{|\mathbf{x_{\oplus}}-\mathbf{x_{\leftmoon}}|^{3}} u^{0}
\label{eq_6EarthMoonSun}
\end{eqnarray}
similar equations for the Sun stand:
\begin{eqnarray}
\frac{dx^{k}_{\odot}}{d\tau}=v^{k}_{\odot} u^{0}  \label{eq_61Sun} \\
\frac{dv^{k}_{\odot}}{d\tau}=-G \frac{M_{\oplus} (x^{k}_{\odot}-x^{k}_{\oplus})}{|\mathbf{x_{\odot}}-\mathbf{x_{\oplus}}|^{3}} u^{0}
-G \frac{M_{\leftmoon} (x^{k}_{\odot}-x^{k}_{\leftmoon})}{|\mathbf{x_{\odot}}-\mathbf{x_{\leftmoon}}|^{3}} u^{0}
\label{eq_6Sun}
\end{eqnarray}
and also for the Moon:
\begin{eqnarray}
\frac{dx^{k}_{\leftmoon}}{d\tau}=v^{k}_{\leftmoon} u^{0}  \label{eq_61Moon} \\
\frac{dv^{k}_{\leftmoon}}{d\tau}=-G \frac{M_{\oplus} (x^{k}_{\leftmoon}-x^{k}_{\oplus})}{|\mathbf{x_{\leftmoon}}-\mathbf{x_{\oplus}}|^{3}} u^{0}
-G \frac{M_{\odot} (x^{k}_{\leftmoon}-x^{k}_{\odot})}{|\mathbf{x_{\leftmoon}}-\mathbf{x_{\odot}}|^{3}} u^{0}
\label{eq_6Moon}
\end{eqnarray} 

The greatest $\phi$ "potential" contribution is the Earth's contribution, which has the following Schwarzschild metric form (see Appendix A):
\begin{equation}
 \phi_{\oplus}=2G\left ( \frac{M_{\oplus}}{r_{\oplus}} \right )
\label{pot_Earth}
\end{equation}
where $ \mathbf{r}_{\oplus}=\mathbf{x}-\mathbf{x}_{\oplus}$. Additionally, 
the satellite geodesics in Schwarzschild perturbed space-time need also be computed. The gravitational potential is therefore: 
\begin{equation}
 \phi=\phi^{\oplus}+\phi^{pert}
\label{pot_total}
\end{equation}
where $\phi^{pert}$ are the additional perturbing potentials produced by the Sun, the Moon and the Earth quadrupole.

The quantities $\phi_{,i}$ are the first spatial derivatives $\partial \phi/\partial x^{i}$, where $x^{i}$ is a function of the proper time $x^{i}=x^{i}\left ( \tau \right )$, so that: 
\begin{equation}
\phi_{,i}=\phi_{,i}^{\oplus}+\phi_{,i}^{pert}
\label{dphi}
\end{equation}

These are the spatial derivatives of the Earth's gravitational potential (central force): 
\begin{equation}
\phi_{,i}^{\oplus}=-2G\left ( \frac{M_{\oplus} \left (x^{i}-x_{\oplus}^{i} \right )}{r_{\oplus}^{3}} \right )
\label{dphi_Earth}
\end{equation}

It should be noted that in this case the reference system is co-moving with the Earth (GRS). 


In order to calculate the satellite world lines 
the time-like geodesic equations from this metric are numerically solved.
To solve the ODE system presented here
a Runge-Kutta method with adaptive step has been implemented.
Our numerical algorithm has
high accuracy  ($10^{-18}$) and multiple precision (40 significant digits for each number). 
This precision will also allow us to consider 
other PN effects as our work advances. Our algorithm is being prepared to provide the precision and accuracy needed for these future additions. 
Notice once more that the effects computed in the present paper consider 
the corresponding terms of Moon or Sun or Earth oblateness or a combination of them.


\pagebreak


\vspace{15cm}
.
\vspace{5cm}
.


\begin{figure}
\center
\includegraphics[width=12cm, height=8cm]{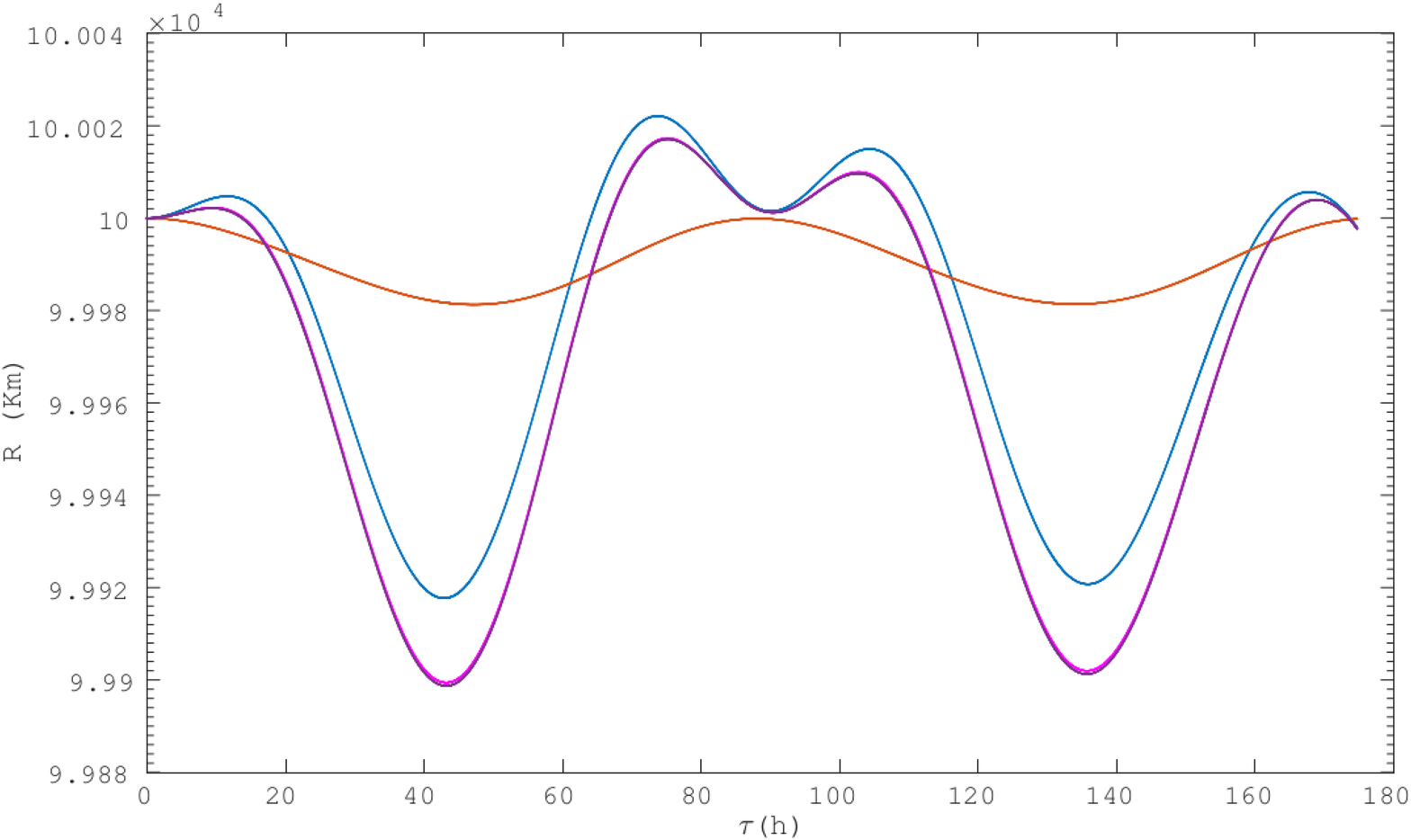} \\
\label{fig_all_efs_ephemerides_100000km}
\end{figure}
\begin{figure}
\center                                                                                                          
\includegraphics[width=12cm, height=8cm]{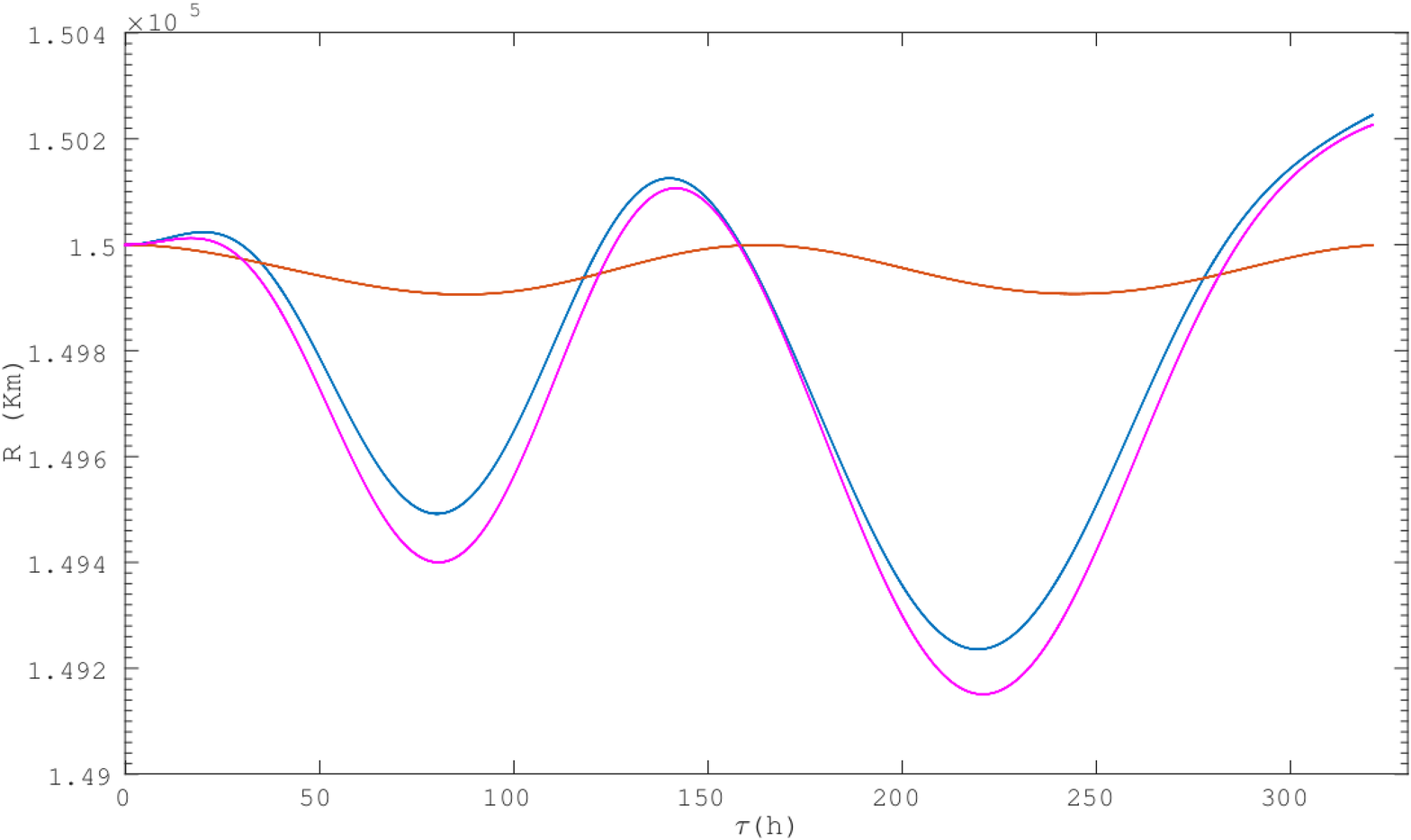} \\
\caption{Radial distance of a satellite versus the proper time (for two orbital periods) at different orbital radius from the geocenter. 
From top to bottom, the considered orbital radius is: $10^{5}$ and $1.5 \times 10^{5}$ in Km. The orbital perturbation effects that 
are considered in the metric are Moon gravitational potential ($blue$), Sun gravitational potential ($red$), 
Moon plus Sun gravitational potential ($magenta$) and, Earth oblateness plus Moon plus Sun gravitational potential ($violet$). 
\label{fig_100000_150000_200000}}
\end{figure}

\vspace{5cm}
.

\begin{figure}
\center
\includegraphics[width=12cm, height=8cm]{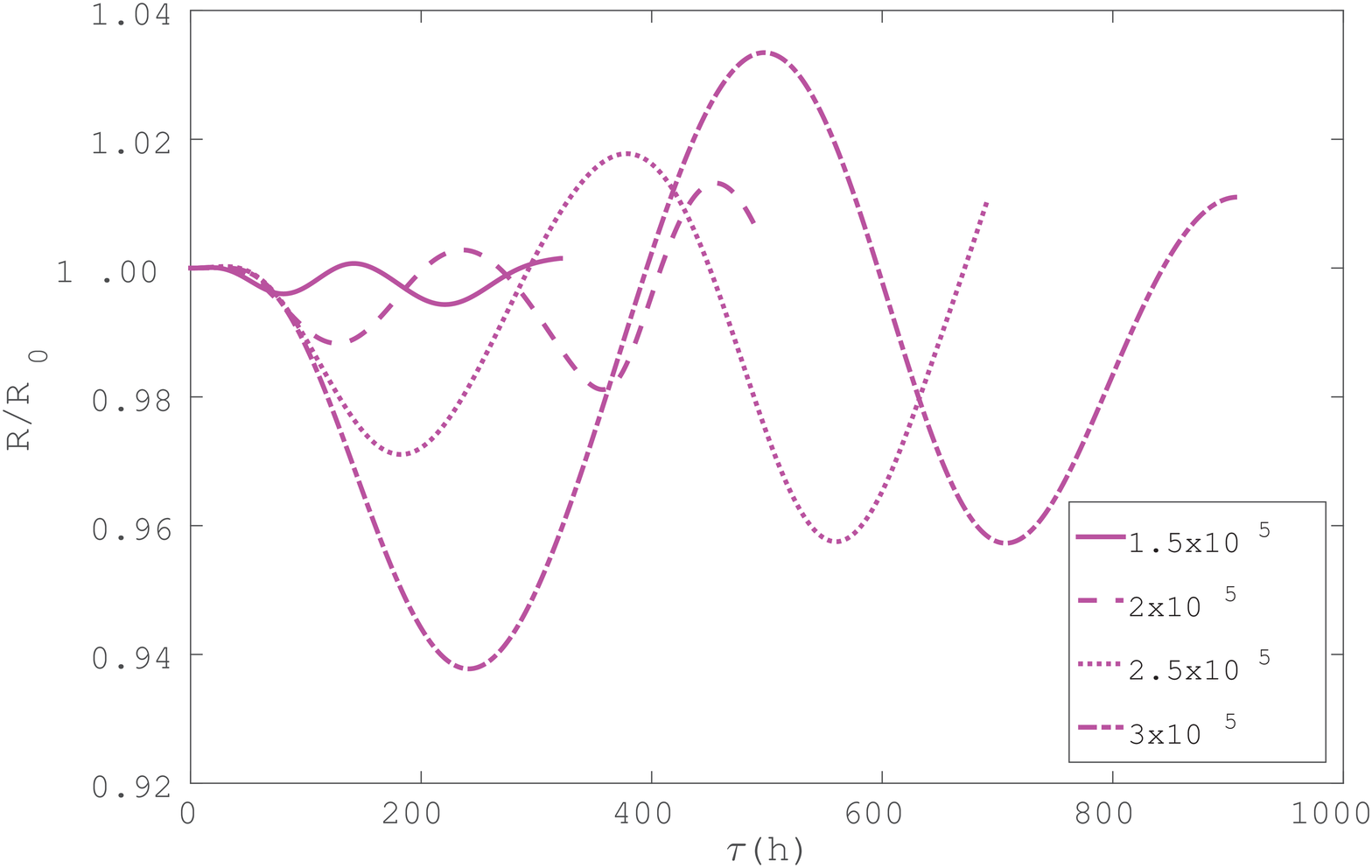} \\ 
\caption{Radial distance $R$, divided by its nominal orbital radius, $R_0$, that is shown in the legend, for a satellite, versus the proper time
(for two orbital periods),
at different orbital radius from the geocenter. 
The orbital perturbation effects that are considered in the metric are Moon plus Sun gravitational potentials ($magenta$).}
\label{fESM_dif_h}         
\end{figure}

\begin{figure}
\centering
\includegraphics[width=12cm, height=8cm]{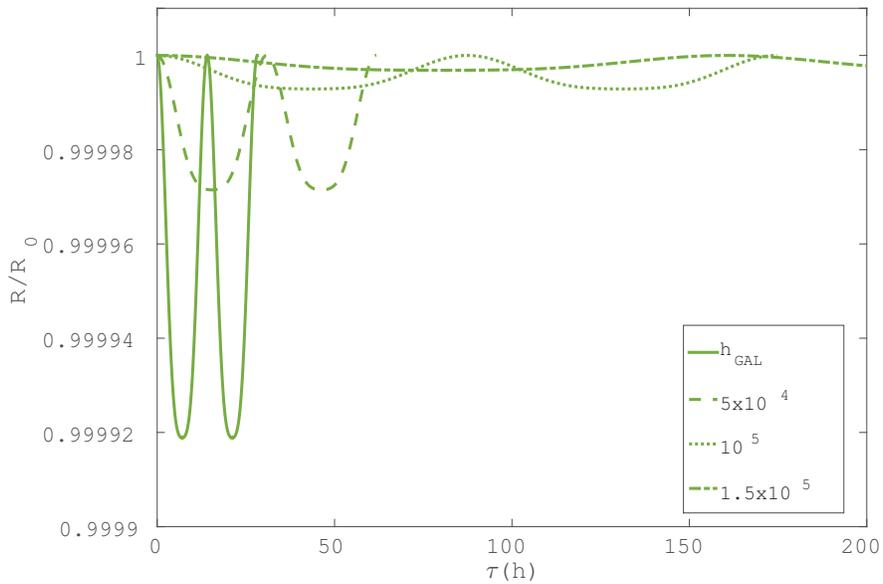}
\caption{Radial distance $R$, divided by its nominal orbital radius, $R_0$, that is shown in the legend, for a satellite, as a function of the proper time (for two orbital periods), 
at different orbital radius from the geocenter. The orbital perturbation effect that is considered in the metric is the Earth oblateness ($green$).
$h_{GAL}$ means the height of a Galileo satelite.}
\label{fig_ESM_ephemerides_dif_h_J2}
\end{figure}

.
\vspace{10cm}
.

\begin{figure}[H]
\center
\label{11h}         
\includegraphics[width=0.35\textwidth]{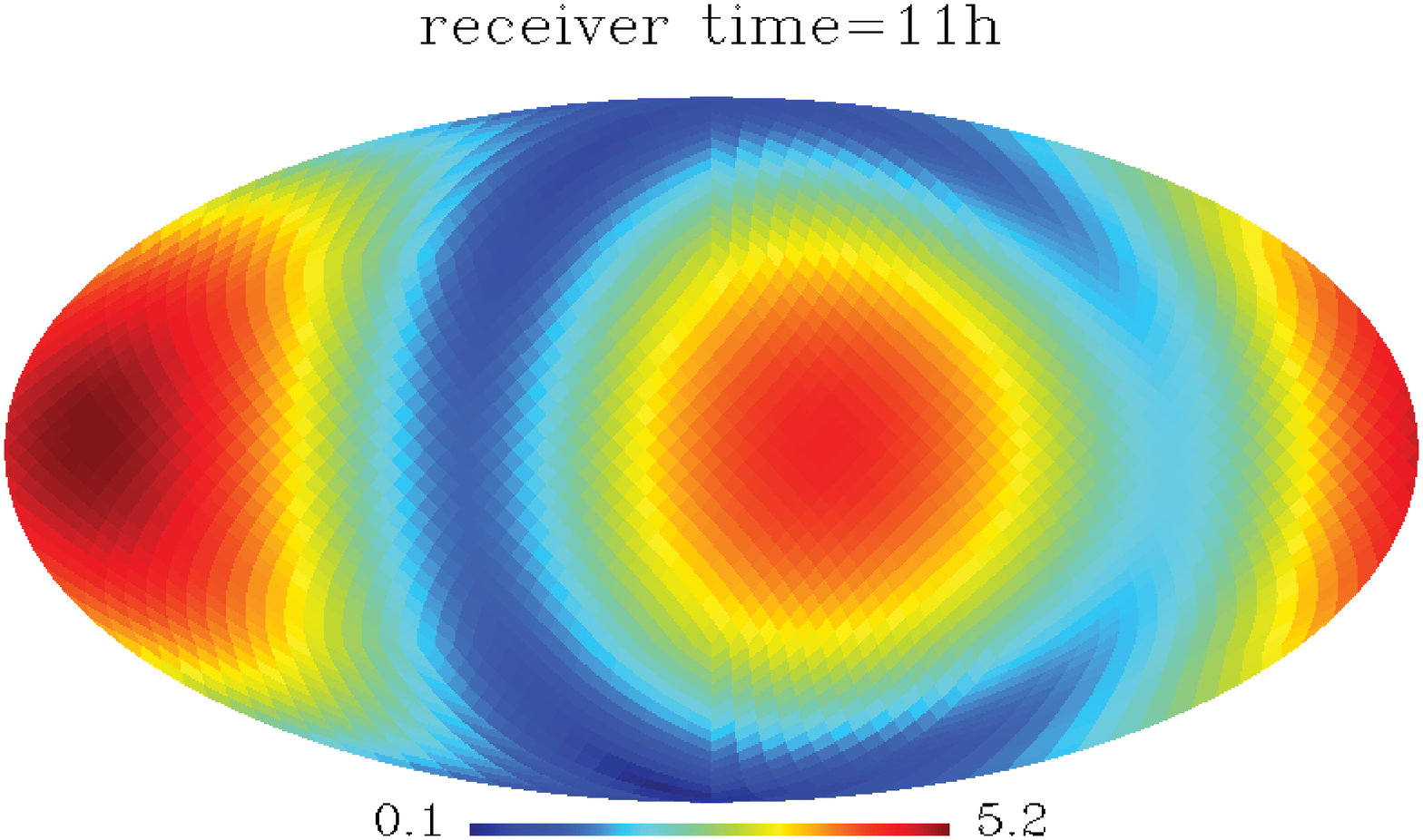}
\end{figure}
\begin{figure}{H}
\center
\label{13h}         
\includegraphics[width=0.35\textwidth]{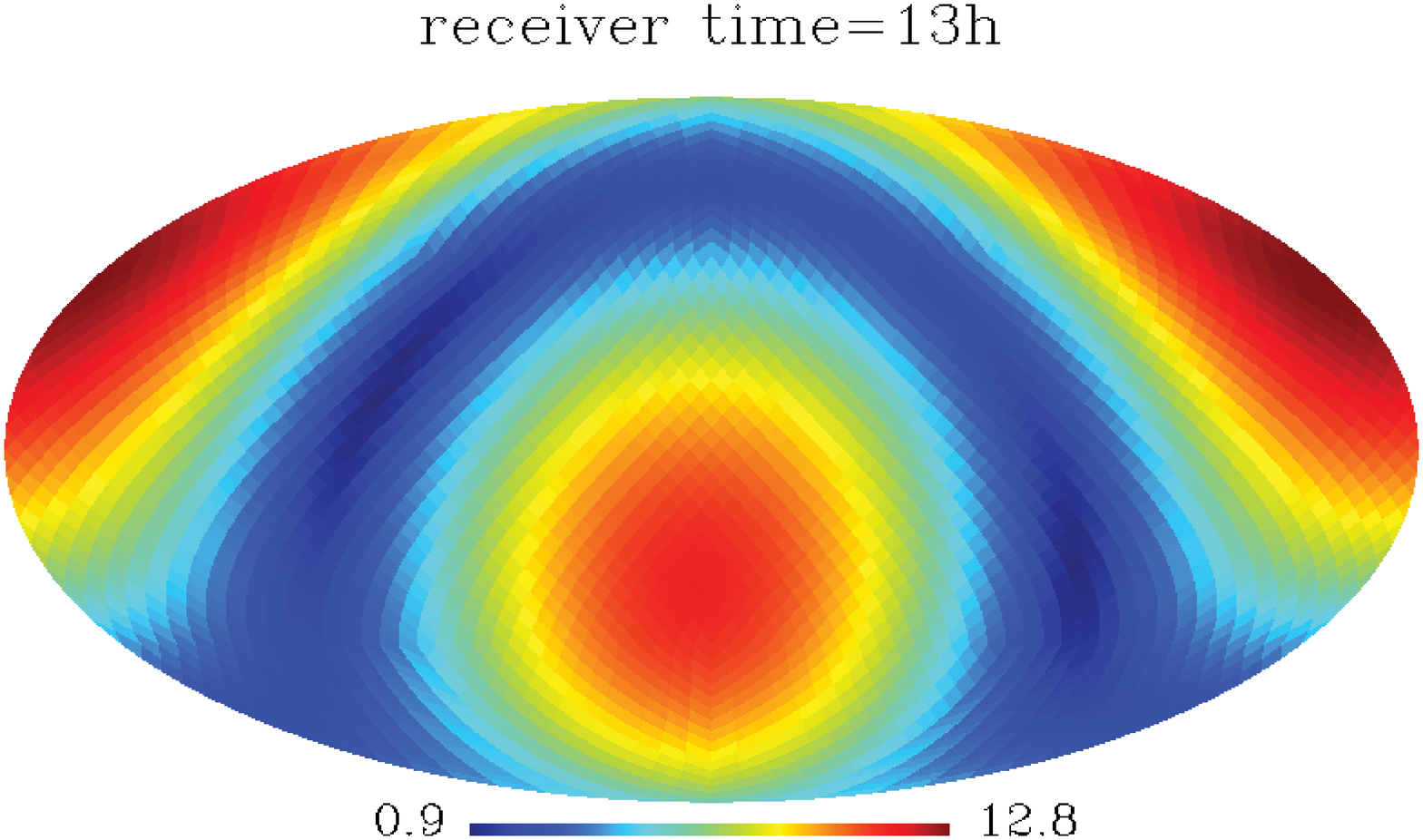}
\end{figure}
\begin{figure}{H}
\center
\label{15h}         
\includegraphics[width=0.35\textwidth]{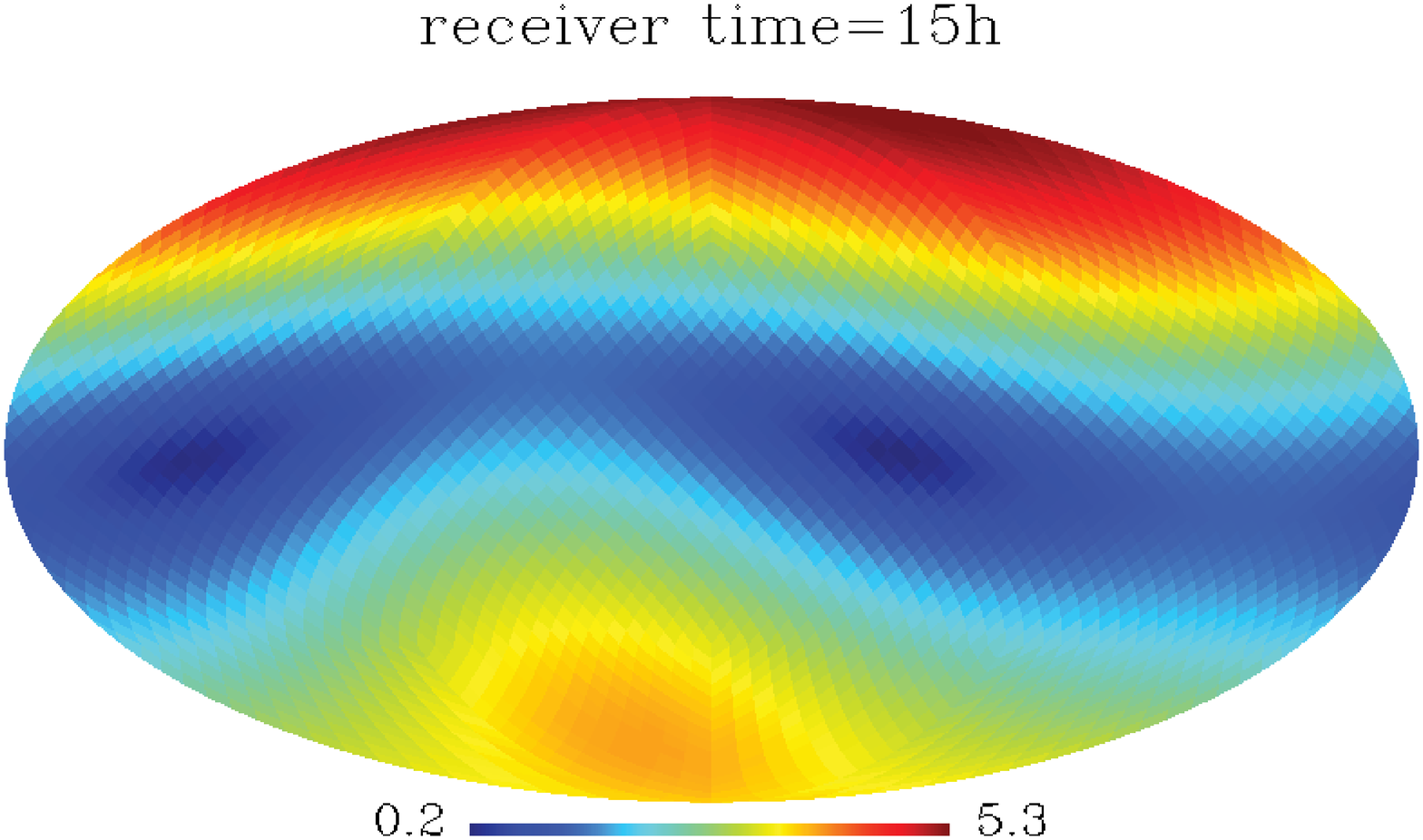}
\end{figure}
\begin{figure}{H}
\center
\label{17h}         
\includegraphics[width=0.35\textwidth]{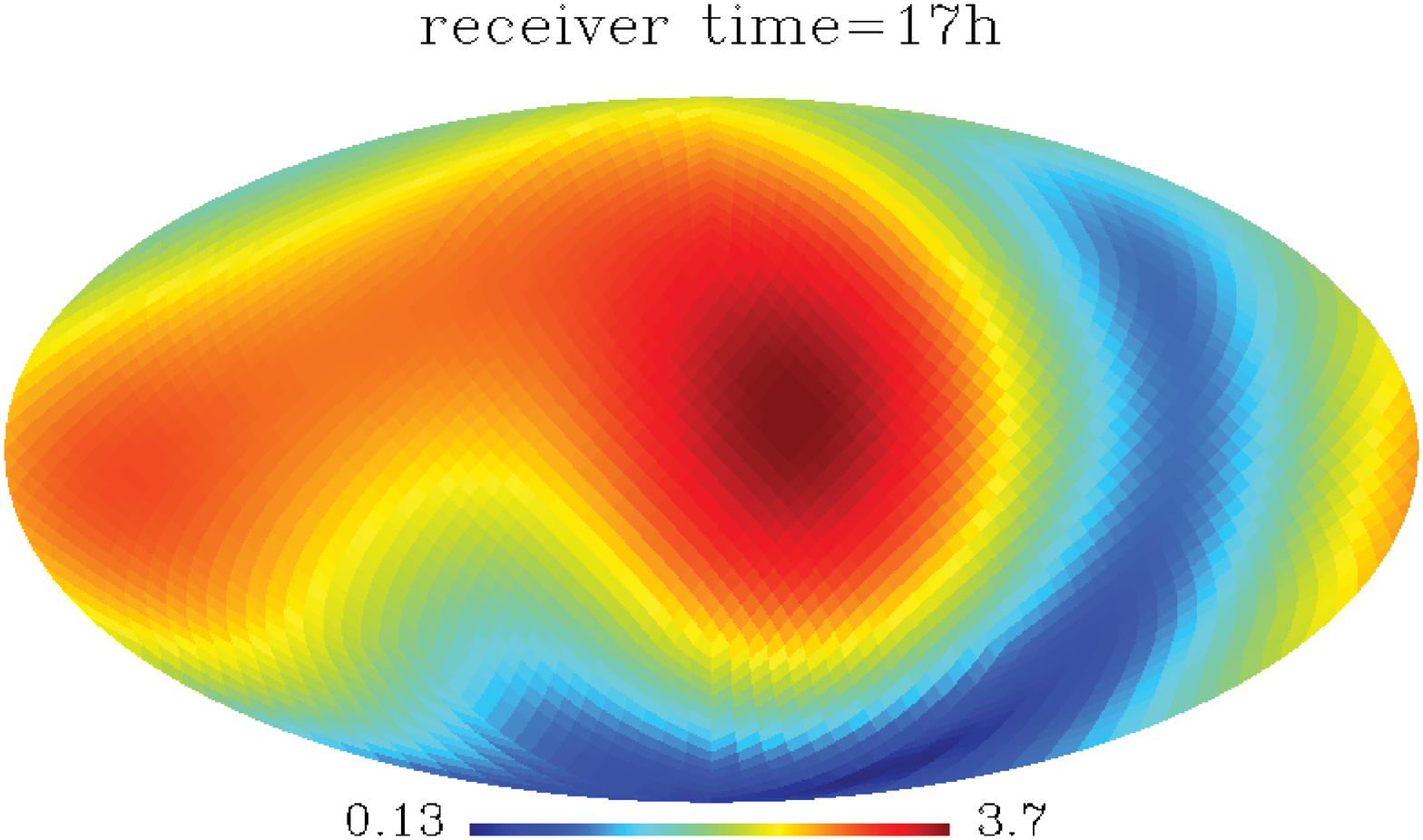}
\caption{HEALPIX mollweide maps of (the error positioning estimator) $\Delta_{d}$ values, in Km, on spheres with Earth radius $6378=R_{\oplus}$ for the Galileo 
satellites 2, 5, 20 and 23 at different user times $t$ (shown in the top of each figure), from $11 \ h$ to $17 \ h$.}
\label{maps_RPS1}
\end{figure}

.
\vspace{15cm}

\begin{figure}{H}
\center
\label{19h}         
\includegraphics[width=0.35\textwidth]{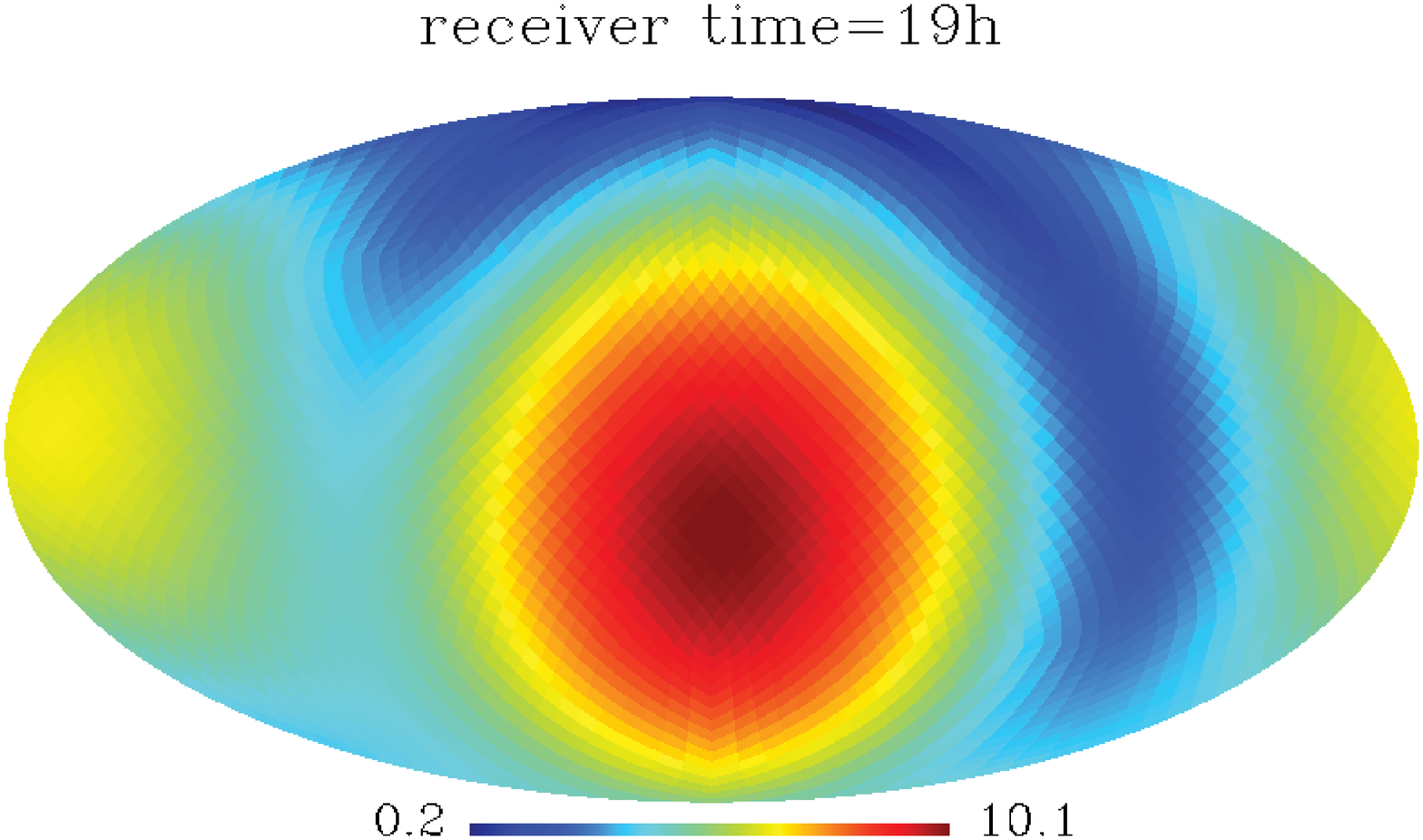}
\end{figure}
\begin{figure}{H}
\center
\label{21h}         
\includegraphics[width=0.35\textwidth]{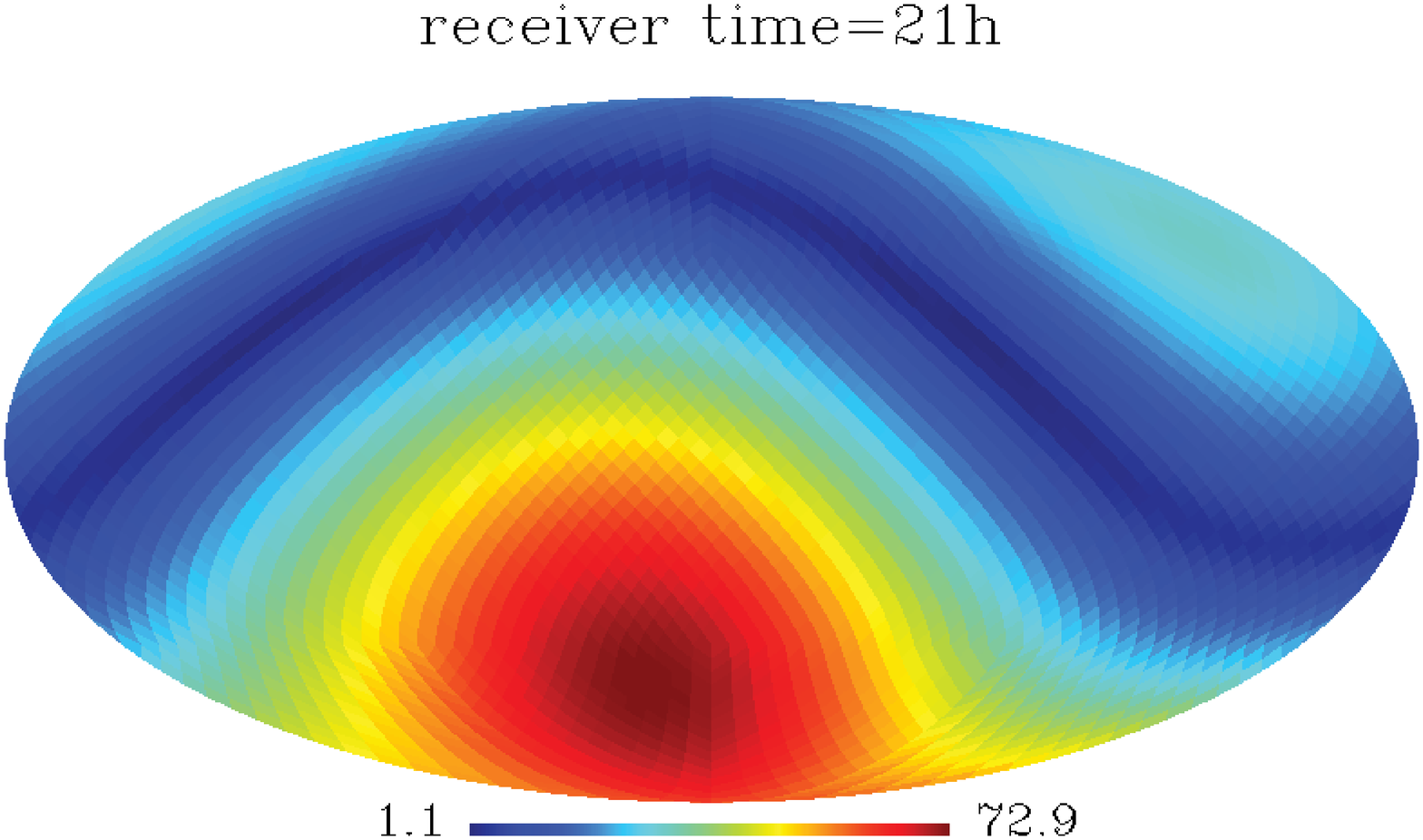}
\end{figure}
\begin{figure}{H}
\center
\label{23h}         
\includegraphics[width=0.35\textwidth]{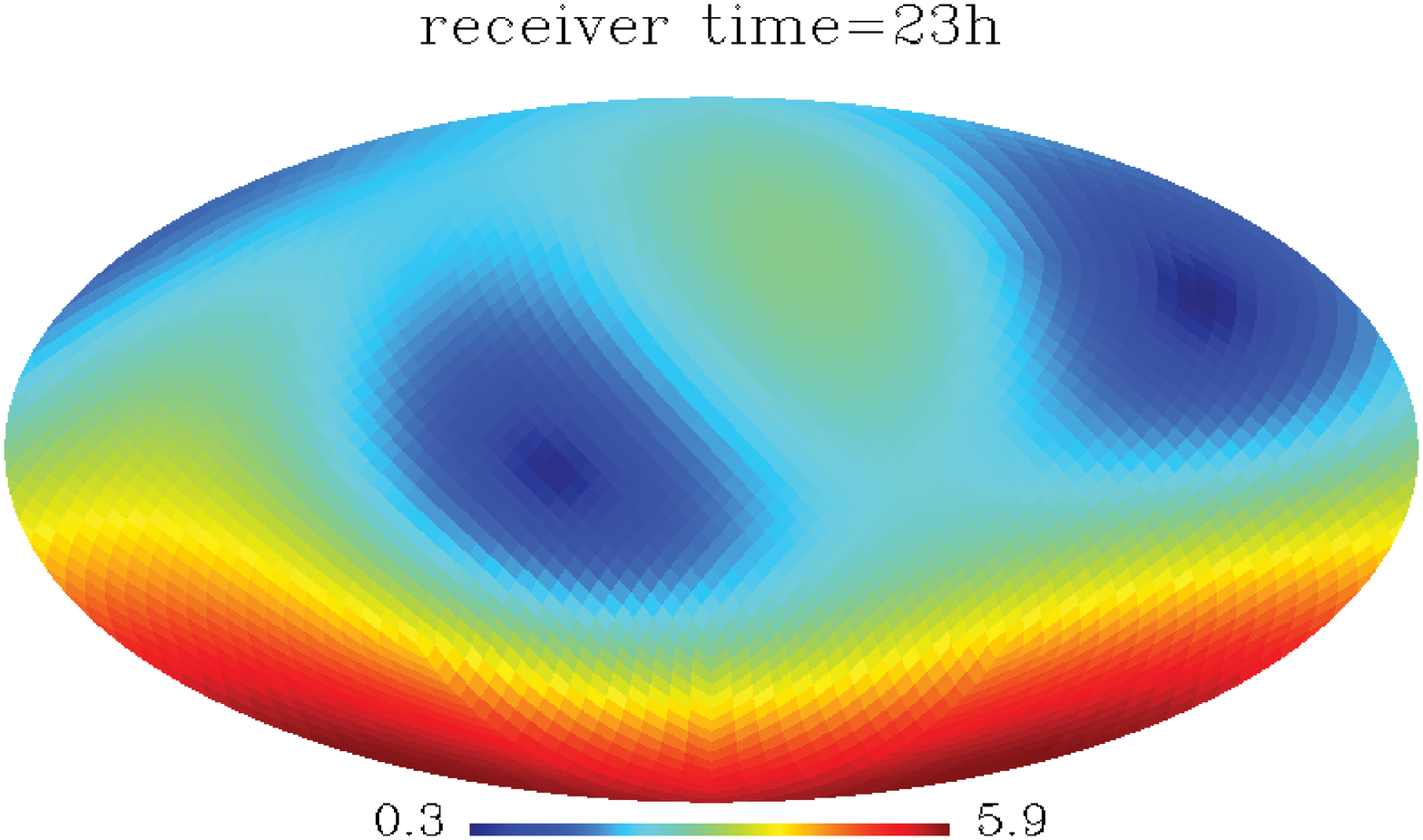} 
\end{figure}
\begin{figure}{H}
\center
\label{25h}         
\includegraphics[width=0.35\textwidth]{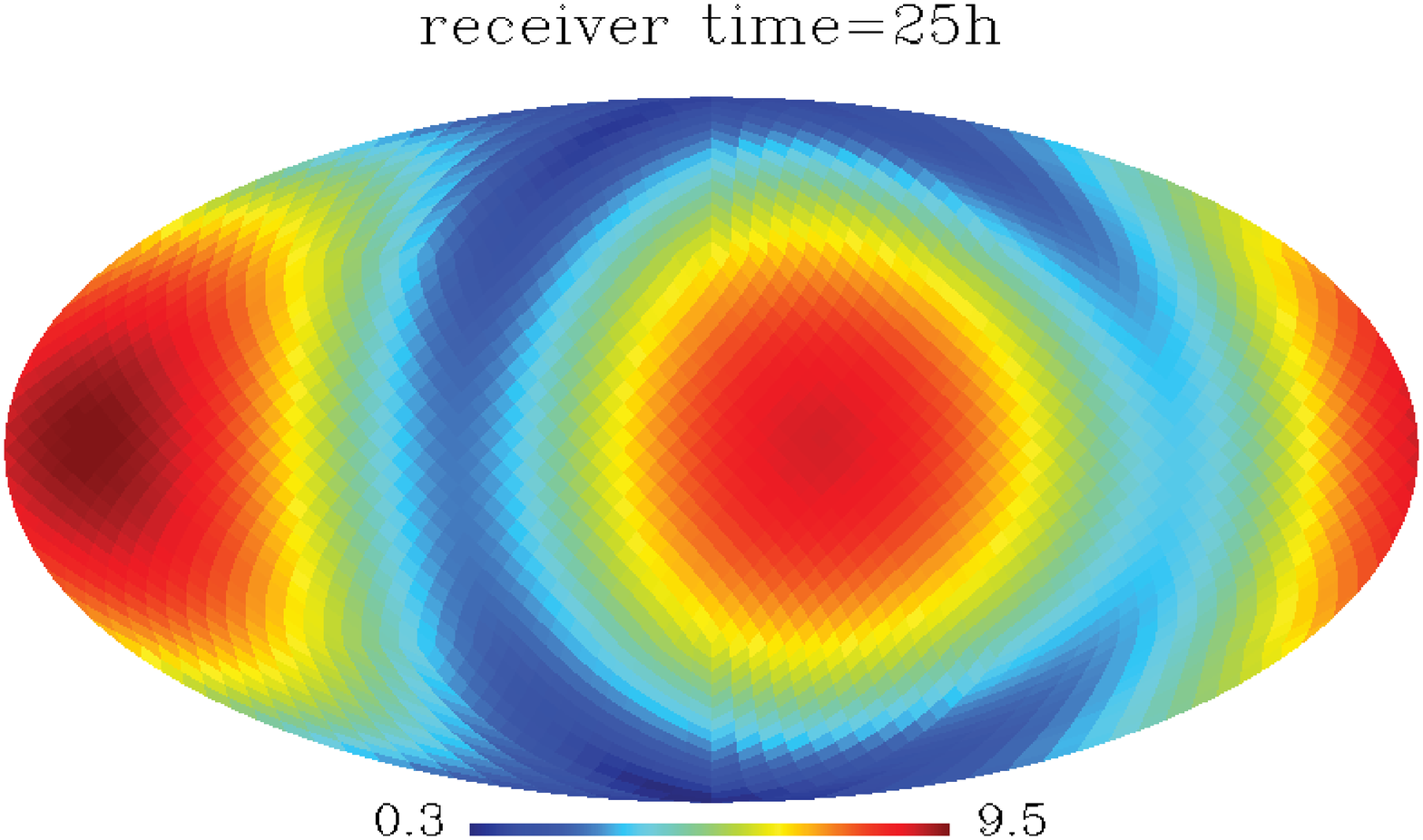}
\caption{HEALPIX mollweide maps of (the error positioning estimator) $\Delta_{d}$ values, in Km, on spheres with Earth radius $6378=R_{\oplus}$ for the Galileo 
satellites 2, 5, 20 and 23 at different user times $t$ (shown in the top of each figure), from $19 \ h$ to $25 \ h$.}
\label{maps_RPS2}
\end{figure}


\begin{figure}{H}
\center
\label{15000}         
\includegraphics[width=0.35\textwidth]{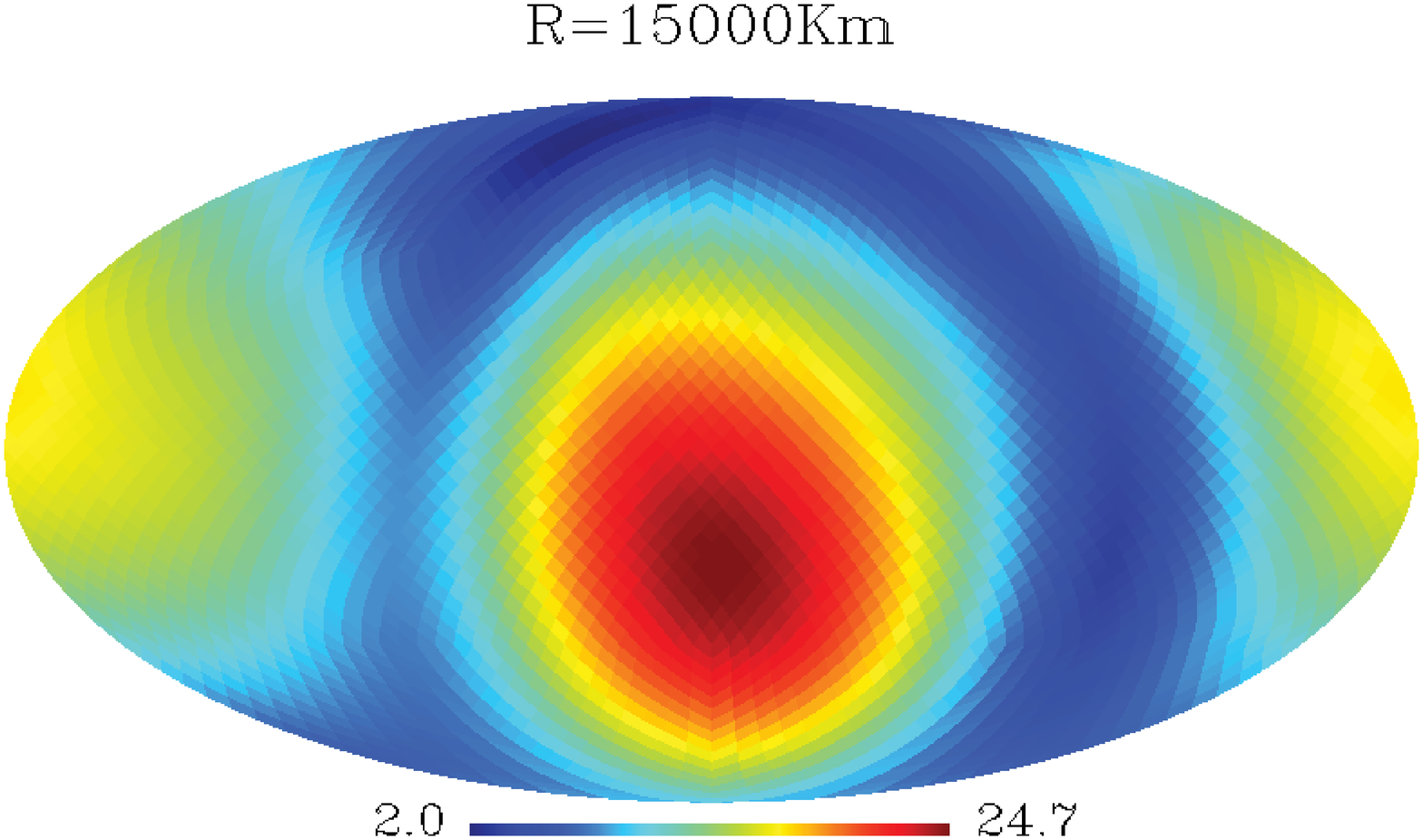}
\end{figure}
\begin{figure}{H}
\center
\label{30000}         
\includegraphics[width=0.35\textwidth]{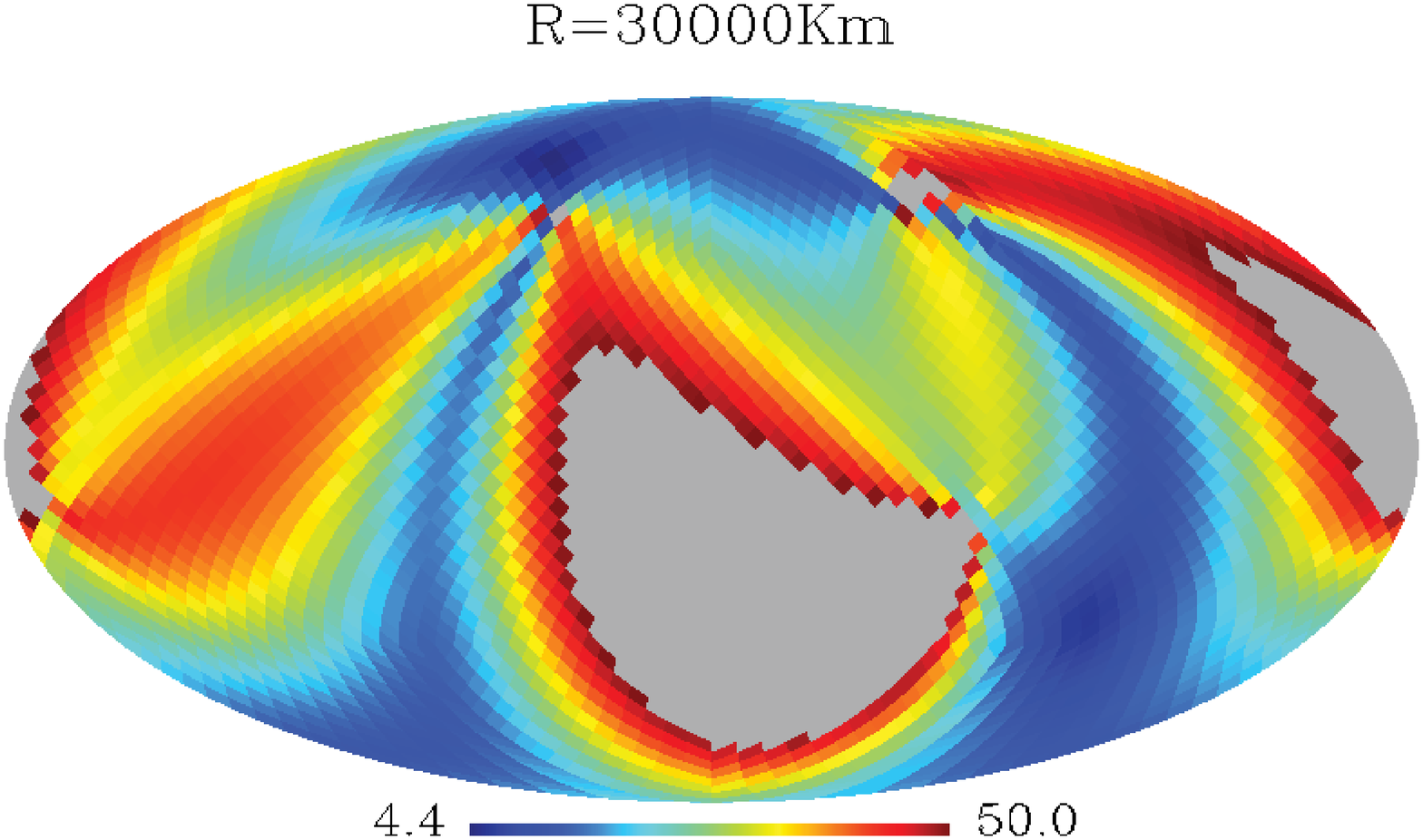}
\end{figure}
\begin{figure}{H}
\center
\label{50000}         
\includegraphics[width=0.35\textwidth]{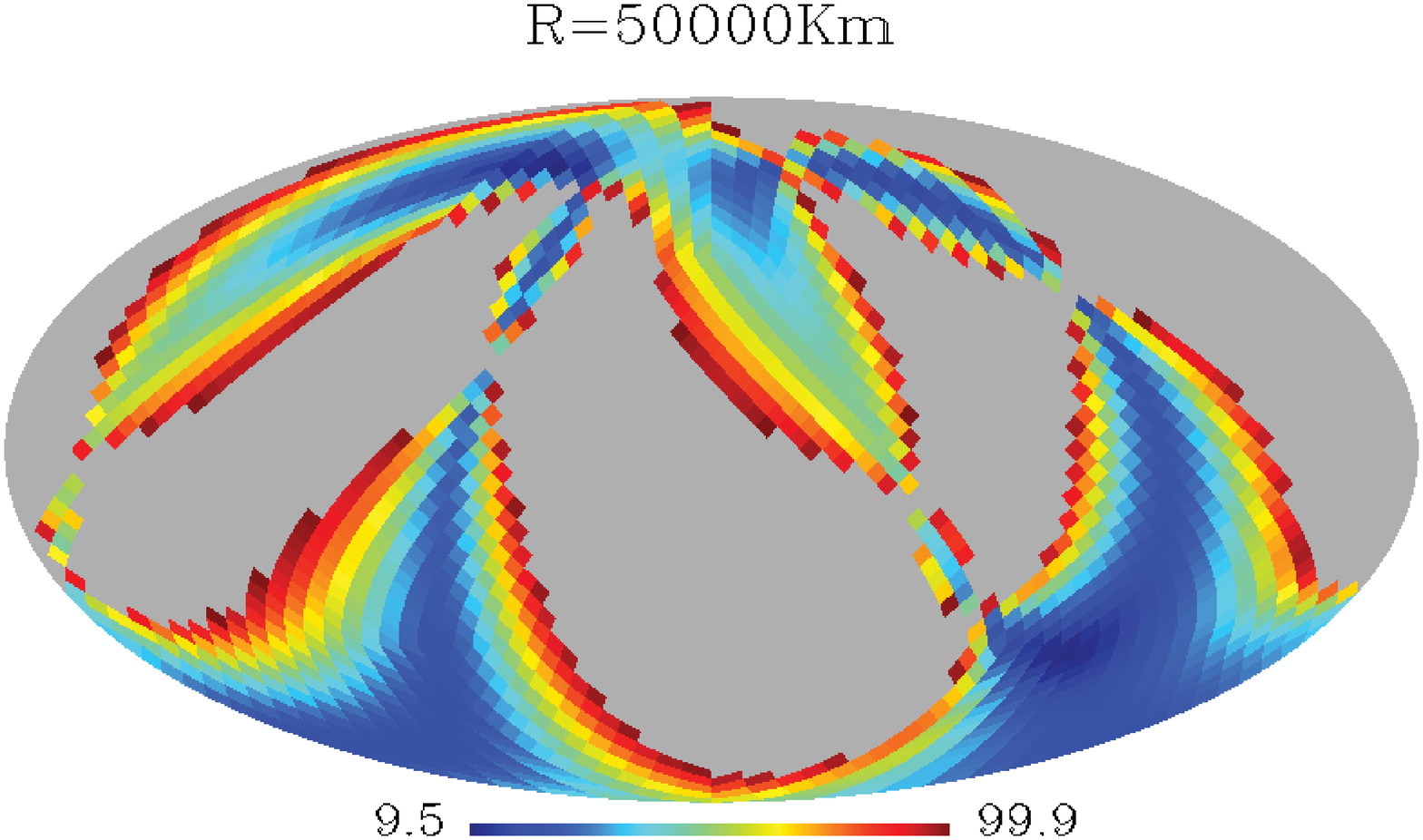}
\caption{HEALPIX mollweide maps of (the error positioning estimator) $\Delta_{d}$ values, in Km, on spheres with different radius 
for the Galileo satellites 2, 5, 20 and 23 at user time $t=19 \ h$. From top to bottom, 
the radius of the spheres in kilometres are $1{.}5 \times 10^{4}$, $3 \times 10^{4}$ and 
$5 \times 10^{4}$ (shown in the top of each figure). Grey coloured pixels are characterized 
by the condition $\Delta_{d} > 50 \ Km$ for the middle subfigure and $\Delta_{d} > 100 \ Km$ for the bottom subfigure.}
\label{maps_RPS3}
\end{figure}

\begin{figure}
\center
\includegraphics[width=0.5\textwidth]{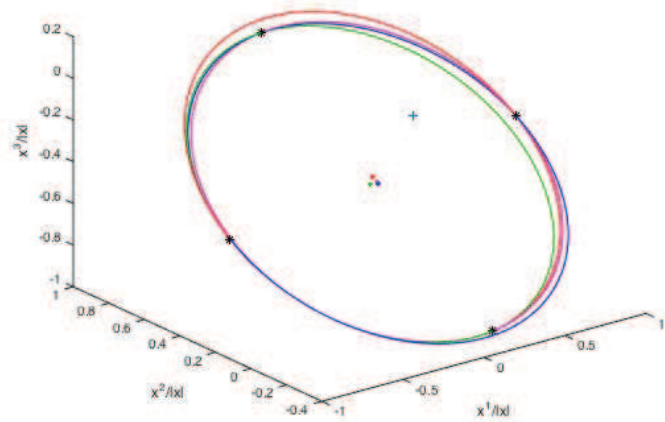}
\caption{Locations of the satellites in the celestial sphere of a user, whose user-satellites configuration corresponds 
to a case of $J \simeq 0$. The considered satellites are 2, 5, 20 and 23 at user time $t=19 \ h$ for a user on a surface of a 
sphere of radius $R=3 \times 10^{4} \ Km$. The satellites considered are represented by four black asterisks, 
the user by a blue cross (the origin of the Cartesian system) and the centres of the circles by red, blue, green and purple dots. 
Four different circles, that pass through three points, represent each possible combination of the three satellites (points).}
\label{celest_sphere_user}
\end{figure}


\label{lastpage}
\end{document}